\documentclass{aa}

\usepackage{graphicx}
\usepackage[varg]{txfonts}
\usepackage[table,xcdraw,dvipsnames]{xcolor}
\usepackage{url}
\usepackage{float}
\usepackage{booktabs}
\usepackage[normalem]{ulem}
\usepackage{mathtools}

\usepackage{hyperref}
\hypersetup{
    colorlinks=true,    
    linkcolor=blue,     
    citecolor=blue,     
    filecolor=magenta,  
    urlcolor=blue       
}
\defcitealias{Anderson2016}{A16}
\defcitealias{Breuval2022}{B22}
\defcitealias{Riess2022}{R22}

\newcommand{\Hcst}{\mbox{$H_{\rm 0}$}\xspace}

\newcommand{\teff}{\mbox{$T_{\rm eff}$}\xspace}
\newcommand{\feh}{\mbox{$\rm{[Fe/H]}$}\xspace}
\newcommand{\mh}{\mbox{$\rm{[M/H]}$}\xspace}
\newcommand{\oh}{\mbox{$\rm{[O/H]}$}\xspace}
\newcommand{\logg}{\mbox{$\log g$}\xspace}

\newcommand{\msun}{\mbox{$\mathrm{M}_{\odot}$}\xspace}

\newcommand{\gaia}{\emph{Gaia}\xspace}

\begin{document} 

    \title{The stellar evolution perspective on the metallicity dependence of classical Cepheid Leavitt laws}

   \author{S. Khan\inst{1}
           \and
           R. I. Anderson\inst{1}
           \and
           S. Ekstr\"om\inst{2}
           \and
           C. Georgy\inst{2}
           \and
           L. Breuval\inst{3}
          }

   \institute{Institute of Physics, \'Ecole Polytechnique F\'ed\'erale de Lausanne (EPFL), Observatoire de Sauverny, 1290 Versoix, Switzerland\\
              \email{saniya.khan@epfl.ch}, \email{richard.anderson@epfl.ch}
    \and
      Department of Astronomy, University of Geneva, Chemin Pegasi 51b, 1290 Versoix, Switzerland
    \and European Space Agency (ESA), ESA Office, Space Telescope Science Institute, 3700 San Martin Drive, Baltimore, MD 21218, USA
             }

   \date{Received \today; accepted MM DD, YYYY}

 
  \abstract{
Metallicity is a key parameter of stellar evolution and significantly affects the mass-luminosity relation of classical Cepheids (henceforth: Cepheids). The impact of metallicity on the Cepheid Leavitt law (period-luminosity relation; henceforth: LL) and, in turn, the Hubble constant ($H_0$), has been the subject of much recent debate. While recent observational results have generally shown a negative intercept-metallicity effect at all wavelengths and for Wesenheit magnitudes, predictions based on different stellar evolution models have differed even in sign.
Here, we present a comprehensive analysis of metallicity effects on Cepheid LLs based on synthetic Cepheid populations computed using Geneva stellar evolution models and the \texttt{SYCLIST} tool. We computed 296 co-eval populations in the age range of $5 - 300$\,Myr for metallicities representative of the Sun, the LMC, and the SMC ($Z \in [0.014, 0.006, 0.002]$). We computed LLs in fourteen optical-to-infrared passbands spanning from \gaia\ $G_{\mathrm{BP}}$ to \textit{JWST}'s F444W and five reddening-free Wesenheit magnitudes. All Cepheid populations take into account distributions of rotation rates and companion stars.
We show excellent agreement between the predicted populations and key observational constraints from the literature, such as a) instability strip (IS) boundaries, b) period distributions, c) LL slopes, d) intrinsic LL dispersion as a function of wavelength and metallicity (match within $0.01$\,mag). This is further strengthened by previously demonstrated excellent agreement between the same models and observed mass-luminosity relations, period-radius relations, and rates of period changes. Our simulations predict a significant LL slope-metallicity dependence ($\beta_\mathrm{M} > 0$) that renders LLs steeper at lower metallicity at all wavelengths. This effect is strongest for shorter passbands where LL slopes are shallower. We point out that observational studies generally support $\beta_\mathrm{M} \ne 0$, albeit with varying degrees of significance. Importantly, $\beta_\mathrm{M} \ne 0$ implies that the intercept-metallicity dependence, $\alpha_\mathrm{M}$, depends on pivot period; an issue not previously considered. Comparison with $\alpha_\mathrm{M}$ measurements in individual passbands reported in the literature yields acceptable agreement on the order of agreement found among different observational studies. The wavelength dependence and magnitude of the disagreement suggests a possible origin in reddening-related systematics. Conversely, we report excellent agreement between our $\alpha_\mathrm{M}=-0.20 \pm 0.03$\,mag\,dex$^{-1}$ and the value determined by the SH0ES distance ladder in the reddening-free $H-$band Wesenheit magnitude $(-0.217 \pm 0.046)$, the currently tightest and conceptually simplest empirical constraint. In summary, we show that Geneva stellar evolution models have high predictive power for Cepheid properties, and that the predicted metallicity effects on the Cepheid LL match the results obtained in parallel with the measurement of the Hubble constant. New evolutionary models extending to super-Solar metallicity and improved astrometry from the fourth \textit{Gaia} data release will allow to further probe these effects.}
   
   \keywords{stars: variables: Cepheids --- stars: distances --- distance scale
               }

   \maketitle
%

\section{Introduction}
\label{sec:intro}

Classical Cepheid variable stars (hereafter Cepheids) have crucial implications both for studies of stellar astrophysics and the extragalactic distance scale. They are evolved intermediate-mass ($M \sim 3$-$8$ \msun) or massive stars ($M \sim 8$-$12$ \msun) that are found in a well-defined region of the Hertzsprung-Russell diagram (HRD), known as the classical instability strip (IS). Up to 9 \msun, these stars may occupy blue loops as they are burning helium in their cores. The range of masses in which this occurs depends on the model input physics, such as metallicity, rotation rate, and convective core overshooting, among others \citep[see, e.g.,][]{Lauterborn1971,Anderson2014,Walmswell2015}.

Cepheids exhibit a tight relation between their pulsation period and luminosity,  
the Leavitt law \citep[][LL]{Leavitt1907,Leavitt1912}, which turns them into standard candles for measuring distances \citep{Hertzsprung1913}. 
The LL is well-characterised from the optical to the infrared. 
Plus, the use of Wesenheit relations has the advantage of mitigating uncertainties and dispersion due to interstellar extinction \citep{Madore1982}, and enabled particular gain in precision, with respect to individual filters. 
The accurate calibration of the Cepheid LL using geometric parallaxes, notably from \textit{Gaia}, anchors the first rung of the distance ladder and, in turn, enables the calibration of the type-Ia supernova (SN) luminosity for the most precise measurement of the local value of \Hcst \citep{Riess2022}.

The influence of metallicity on Cepheid distances has prominently featured in the recent literature on Cepheids and \Hcst measurements \citep[e.g.][]{Breuval2022,Riess2022,Trentin2024}. Within the distance ladder, Cepheids are split between those pertaining to nearby galaxies, for instance the Milky Way (MW) or Magellanic Clouds (MCs), for which distances can be measured geometrically; and more distant ones in SN-host galaxies, in which case distances are obtained by fitting LLs. These two samples are typically referred to as the \textit{anchor} and \textit{host} sets. Chemical composition varies from star to star, and both the anchor set of Cepheids and the host set of Cepheids span a range of metallicities. 
Chemical abundances of Cepheids in the MW and in the MCs can be measured using high-resolution spectroscopy. Recent measurements suggest $\feh = -0.409 \pm 0.003$ and a dispersion of 0.076 dex for the LMC \citep{Romaniello2022}, and $\feh = -0.785 \pm 0.012$ and a dispersion of 0.082 dex for the SMC \citep{Breuval2024}.
The MW alone covers the broadest range of metallicities, $-1.1 < \rm{[Fe/H]} < +0.3$ \citep{Trentin2023}, which correlates with distance due to the Galactic metallicity gradient \citep[e.g.,][]{Andrievsky2002,Lemasle2008,Genovali2014}. 
Individual star spectroscopy is however not feasible for Cepheids in SN-host galaxies. In that case, metallicities, or more specifically oxygen abundances, are estimated based on the galactocentric location and metallicity gradients estimated using HII regions \citep[e.g.,][]{Bresolin2011}. Subsequent checks are done to ensure that oxygen abundance gradients in the MW inferred by spectra of Cepheids and from HII regions are consistent with one another \citep[see Fig. C1 in][henceforth: \citetalias{Riess2022}]{Riess2022}. Metallicities of Cepheids in SN-host galaxies are contained within a relatively small range of metallicities present in the anchors \citepalias[cf. Fig. 21 in][]{Riess2022}. A significant bias in \Hcst would require both a systematic difference between hosts vs anchors and a significant metallicity effect on Cepheid luminosity, which clearly is not the case. However, a detailed quantification of metallicity effects serves to both improving precision on Cepheid distances (thus, on \Hcst) and elucidating stellar models.

We adopted the below nomenclature to describe the LL's metallicity effects, cf. Appendix\,\ref{app:PLZ} for more details and Table\,\ref{tab:terminology}, which compares our nomenclature to the recent literature. 
Metallicity effects are considered for both the LL intercept $\alpha$ and the slope $\beta$ as follows:
\begin{align}
   M  &=  \alpha + \beta \cdot \log{(P/P_0)} \label{eq:pl_general} \\
      &= \alpha_0 + \alpha_{\mathrm{Z}}\cdot Z + \left( \beta_0 + \beta_{\mathrm{Z}}\cdot \mathrm{Z} \right)\cdot \log{(P/P_0)} \\
      &= \alpha_0 + \alpha_{\mathrm{M}}\cdot \mh + \left( \beta_0 + \beta_{\mathrm{M}}\cdot \mh \right)\cdot \log{(P/P_0)}
      \label{eq:plz_full}
\end{align}
where $\mh=\log{(Z/Z_{\rm \odot})}$ and $\log$ is shorthand for $\log_{10}$. Subscripts $_\mathrm{Z}$ and $_\mathrm{M}$ indicate that the effect is assessed by overall metallicity (rather than a specific abundance) as defined in the models. Subscript $_0$ readily identifies fiducial quantities, and $P_{\rm 0}$ the pivot period. In comparing our model predictions to observations, we  assumed equivalence between the luminosity dependence on overall metallicity \mh and iron abundance \feh. A metallicity-dependent LL slope, with $\beta_{\rm M} \neq 0$, implies that the metallicity dependence of the intercept,  $\alpha_{\rm M}$, depends on $P_{\rm 0}$.

Empirical studies of the LL metallicity dependence have greatly improved over the last several years. \citet{Gieren2018} adopted homogeneously-measured distances from a Baade-Wesselink analysis to our Galaxy, LMC, and SMC. \citet[henceforth: \citetalias{Breuval2022}]{Breuval2022} studied the Cepheid LL in the MW and the MCs (with a slope fixed to that of the LMC), using distances based on \gaia EDR3 parallaxes and eclipsing binaries, respectively. In their most recent study, \citet{Breuval2024} adopted \textit{HST} photometry for SMC Cepheids. Other recent works have used the MW alone, together with \gaia-based distances, to quantify the metallicity effect \citep{Ripepi2022,Bhardwaj2023,Bhardwaj2024,Trentin2024}. All these studies led to the finding that $\alpha_{\rm Fe}$ is negative at all wavelengths, with notable differences in the scale.

On the modelling side, {different stellar evolution codes and their respective input physics led to different results. Stellar evolution models from \citet{Ekstrom2012,Georgy2013} combined with a linear non-adiabatic radial pulsation analysis from \citet[][hereafter \citetalias{Anderson2016}]{Anderson2016} have been tested empirically (relations between period and luminosity, radius, effective temperature, and period change rates) and shown very good agreement regarding the location of the IS in $\log \teff$-$\log L$ space \citep[see Fig. 12 in][]{Espinoza-Arancibia2024}. The study by \citetalias{Anderson2016} had already suggested a metallicity effect on the LL slope using single-star models (as illustrated in their Fig. 15). A comparison of those models, with $\log P_{\rm 0}=0.0$, with observations in \citetalias{Breuval2022} shows a good agreement, but potential differences arising due to the choice of pivot period have not been considered. Theoretical studies based on updated versions of the Stellingwerf code \citep{Bono1994} reported negative $\alpha_{\rm M}$ for Wesenheit magnitudes and positive $\alpha_{\rm M}$ for individual passbands \citep{DeSomma2020,DeSomma2022,DeSomma2024}. 
The strong dependence of the IS boundaries on metallicity in these models \citep{DeSomma2022} may explain this behaviour, since Wesenheit magnitudes combine a single-band magnitude with a colour term. 
For reference, \citetalias{Anderson2016} reported a weak dependence of IS boundaries on metallicity and negative values of $\alpha_{\rm M}$ both for individual passbands and Wesenheit magnitudes. 

The present article aims to conclusively explore the metallicity dependence of Cepheid luminosity based on the models and instability analysis presented in \citetalias{Anderson2016}. To this end, we considered 
a large number of photometric passbands and Wesenheit relations, notably including the \gaia and \textit{JWST} photometric systems. We computed synthetic populations of Cepheids using a modified version of the \texttt{SYCLIST} program \citep{Georgy2014} to improve realism and assess sampling effects across the IS. The article is structured as follows. 
Section \ref{sec:method} presents our methodology, including a short description of the models and simulations used and the transcription of modelled quantities to observational properties of Cepheids, such as magnitudes and periods. Section \ref{sec:properties} compares our predictions for the period distribution and sampling of the IS with empirical results, and introduces our LL fits based on both \texttt{SYCLIST} synthetic populations and IS edges. Throughout the paper, we compared the theoretical predictions to observations. Results concerning the LL metallicity dependence are presented in Sect. \ref{sec:LL_metallicity}. The final Sect. \ref{sec:conclusions} summarises our results and presents our conclusions.

\section{Method}
\label{sec:method}

\begin{figure*}
    \centering
    \includegraphics[width=\hsize]{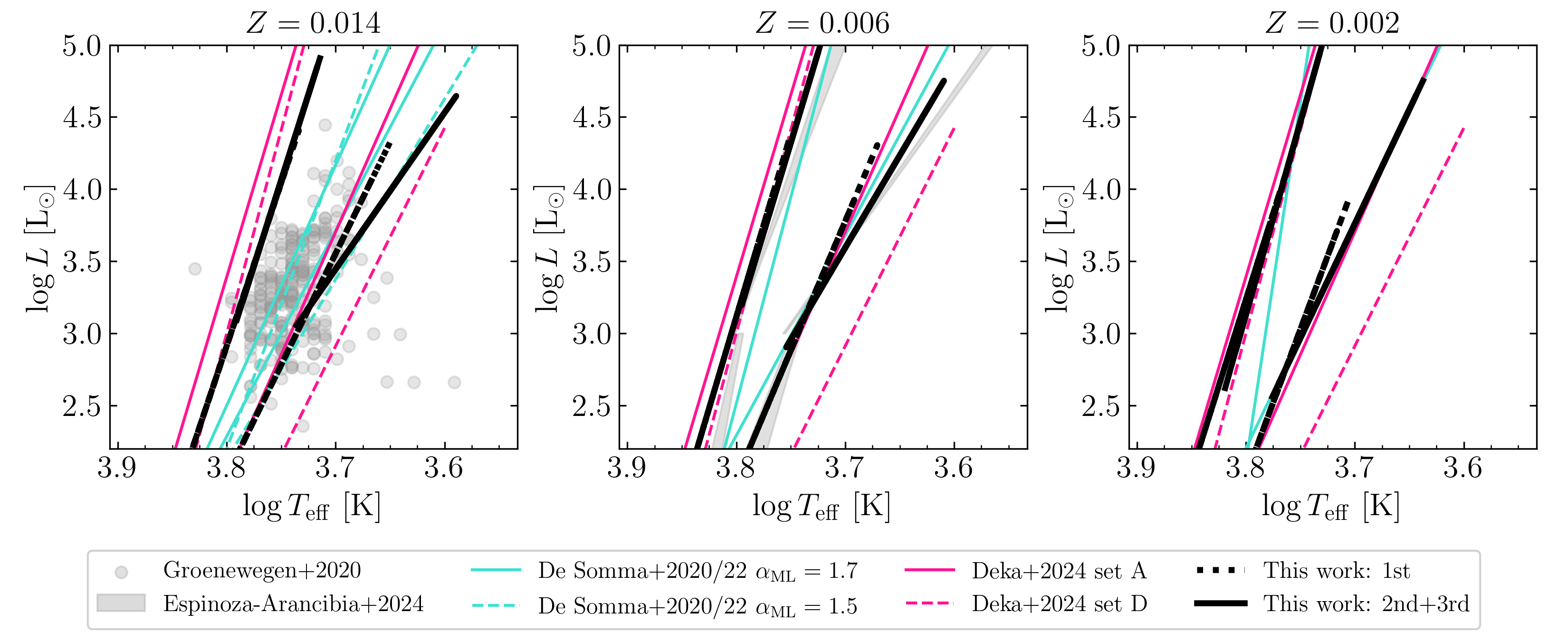}
	\caption{Hertzsprung-Russell diagrams for metallicities $Z=0.014$ (Solar, left), $Z=0.006$ (LMC, middle), and $Z=0.002$ (SMC, right). Solid black lines show predicted IS boundaries from  \citetalias{Anderson2016} for the second and third crossings averaged together; dashed black lines show the first crossing. Solid and dashed cyan lines illustrate predicted boundaries from \citet[their case A called ``canonical'' mixing-length relation; note the difference in $Z$ values: $0.020$, $0.008$, $0.004$]{DeSomma2020,DeSomma2022}, for $\alpha_{\rm ML}=1.7$ and 1.5, respectively. Solid and dashed pink lines show predicted sets A (simple convective model) and D (added radiative cooling, turbulent pressure \& flux) from \citet{Deka2024}, which estimate the envelopes of instability strips across a large range of metallicities (\mh $\in$ \{0.00, $-$0.34, $-$0.75\}, i.e., $Z \in$ \{0.013, 0.006, 0.002\}) and are shown in all panels.
    Observational estimates of 265 fundamental-mode Galactic classical Cepheids from \citet[][]{Groenewegen2020} are shown in the left panel. Empirically-derived IS boundaries for the LMC from \citet{Espinoza-Arancibia2024} are shown as grey regions in the middle panel.}
	\label{fig:IS}
\end{figure*}

We base this work primarily on synthetic Cepheid populations computed using \texttt{SYCLIST}, that is based on Geneva stellar evolution models for which \citetalias{Anderson2016} performed the pulsational instability analysis. We first review the basics of the models (Sect. \ref{sec:models_a16}) and consider the adequacy of the IS boundaries (Sect. \ref{sec:IS}) before presenting the synthetic populations (Sect. \ref{sec:syclist}).

\subsection{Description of the models}
\label{sec:models_a16}
We relied on Geneva stellar evolution models presented in \citet{Ekstrom2012} and \citet{Georgy2013}. The properties of Cepheids predicted by these models, notably with respect to the effects of rotation, were investigated by \citet{Anderson2014} and \citetalias{Anderson2016}.
The grid of evolutionary tracks was computed using the following initial masses ($\mathcal{M}$): 1.7, 2, 2.5, 3, 4, 5, 7, 9, 12, and 15 \msun, and three metallicities: $Z=0.014$, 0.006, and 0.002 (or, equivalently, $\mh=0.000$, $-0.368$, and $-0.845$), corresponding to the chemical abundances of young and intermediate-age populations in the MW, the LMC, and the SMC, respectively. The initial helium mass fraction is determined assuming a linear chemical enrichment law, $Y(Z)=Y_{\rm P} + \frac{\Delta Y}{\Delta Z} Z$, with $Y_{\rm P}=0.2484$ the primordial He abundance \citep{Cyburt2003}, and $\Delta Y / \Delta Z = 1.257$ \citep[see Sect. 3.1 in][]{Georgy2013}. For the Solar mixture, we thus have $Z=0.014$, $Y=0.266$, and $X=0.720$ (using $X = 1 - Y - Z$) as initial abundances of metals, He, and H. Our models assume the mixture of heavy elements from \citet{Asplund2005}.

The mixing-length parameter is taken equal to $\alpha_{\rm ML} = 1.6$. Convective core overshooting is included at the level of $0.10 \, H_{\rm P}$ during the H- and He-burning phases, where $H_{\rm P}$ is the pressure scale height at the edge of the convective core as given by the Schwarzschild criterion. 
The transport of angular momentum due to rotation is implemented via the advecto-diffusive scheme using a horizontal  diffusion coefficient $D_{\rm h}$ \citep{Zahn1992} and a shear coefficient $D_{\rm shear}$ \citep{Maeder1997}.
Three initial rotation rates were used: $\omega = \Omega/\Omega_{\rm crit}=0.0$, 0.5, and 0.9, corresponding to no, intermediate (typical), and fast rotation. $\Omega_{\rm crit}$ refers to the hydrostatic definition of the critical angular velocity, i.e. $\Omega_\text{crit}=\sqrt{\frac{8}{27}\frac{G\mathcal{M}}{R^3_\text{p,crit}}}$, where $R_\text{p,crit}$ is the polar radius when the star is at the critical rotation. Radiative mass loss is implemented using different prescriptions depending on the initial mass and the evolutionary stage \citepalias[see Table 1 in][]{Anderson2016}, and mechanical mass loss occurs when the surface velocity of the star reaches the critical velocity, which is more likely to affect fast rotators on the higher-mass end during the main sequence (MS).

Further details about the models and their calibration are provided in \citet[][Sect. 2]{Ekstrom2012} and \citet[][Sects. 2 and 3]{Georgy2013}. Importantly, we stress that model calibration was based on MS stars only.

\subsection{Position of the instability strip}
\label{sec:IS}
The location of the IS boundaries is a key element for predicting Cepheid properties based on stellar models and can differ substantially among different models. Some discussion of the adequacy of the IS boundaries used here is thus warranted.

\citetalias{Anderson2016} computed IS boundaries and pulsation periods for fundamental mode and first overtone Cepheids based on the Geneva models using a linear non-adiabatic radial pulsation analysis. The method was described in \citet{Saio1983} and  supplemented with the effects of pulsation-convection coupling using time-dependent convection theory \citep{Unno1967,Grigahcene2005}, which allows to determine red IS edges while also considering the impact of convection on the blue IS edge. Including the coupling significantly improves the agreement with observations \citepalias[see Fig. 2 in][]{Anderson2016}. Further details are provided in Sects. 2 and 3.1 of \citetalias{Anderson2016}.
For the present work, we considered exclusively fundamental mode Cepheids, which are most relevant to the distance ladder. 
We considered IS boundaries obtained by averaging all three rotation rates ($\omega = 0.0, 0.5, 0.9$) and combined second and third crossings. First crossings were kept separate. 

\begin{figure*}
    \centering
    \includegraphics[width=\hsize]{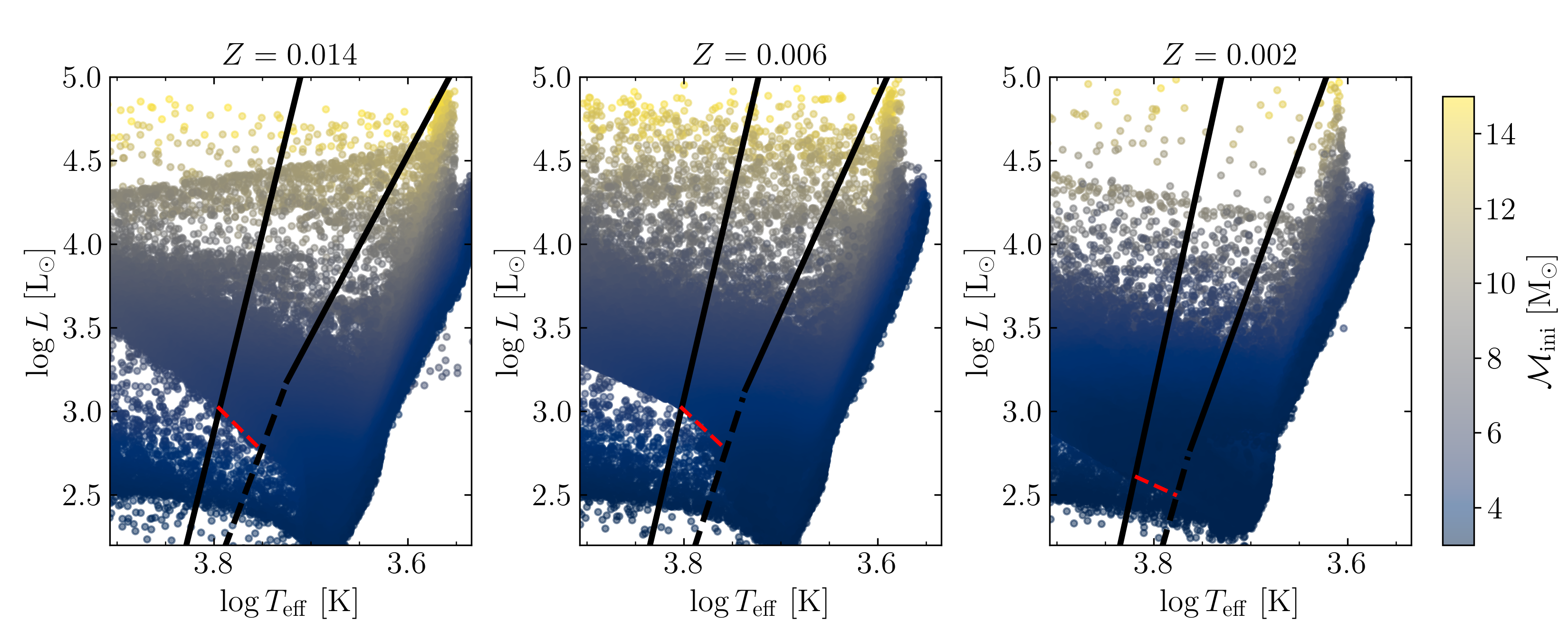}
	\caption{Hertzsprung-Russell diagrams for all 296 \texttt{SYCLIST} cluster simulations at each metallicity considered: $Z=0.014$ (left), $Z=0.006$ (middle), and $Z=0.002$ (right). The colour scale indicates the initial mass of the star, from the least (blue) to the most massive ones (yellow). Overlaid in black are the IS boundaries determined by \citetalias{Anderson2016} for second and third crossings averaged together (solid), and first crossing (dashed). The red dashed lines show the cut applied to exclude most of the first crossing stars.}
	\label{fig:SYCLIST_HRD}
\end{figure*}

Figure \ref{fig:IS} compares predicted IS boundaries from \citetalias{Anderson2016} both to other models and empirical results. Specifically, it compares our predictions to 265 Galactic classical Cepheids from \citet{Groenewegen2020}, for which \teff and bolometric $L$ are derived by fitting model atmospheres to the SEDs and using distance and reddening values from the literature. They fall well within our predicted IS, apart from several low-luminosity Cepheids that appear significantly too cool. \citet[see their Fig.~3]{Javanmardi2021} and \citet[][Fig. 3]{Trahin2021} previously also showed excellent agreement between the \citetalias{Anderson2016} IS boundaries and the location of 28 and 63 Cepheids, respectively, analysed in great detail using the SPIPS algorithm \citep{Merand2015}. Unfortunately, a dearth of high-luminosity Galactic Cepheids near the red IS edge leaves this IS edge poorly constrained empirically. However, the changing slope of the red IS boundary is well documented in the LMC, where our predictions closely match empirically derived IS boundaries \citep{Espinoza-Arancibia2024}. 
We note that observed Cepheid populations generally span a range of metallicities, whereas model predictions are mono-metallic.

On the theory side, \citet{DeSomma2020,DeSomma2022} computed IS boundaries based on non-linear hydrodynamical models assuming three different mass-luminosity relations during the Cepheid stage, whereby case A is computed for ``canonical'' models, i.e., without convective core overshooting, rotation, and mass loss. Cases B \& C consider ad hoc increases in luminosity at fixed mass by adding $\Delta\log{L}=0.2$ and $0.4$\,dex to the canonical model, respectively.
\citet{DeSomma2020,DeSomma2022} further considered sensitivity to the mixing-length parameter $\alpha_{\rm ML}$. Figure \ref{fig:IS} shows their IS edges for canonical models with $\alpha_{\rm ML} = 1.7$, as well as $\alpha_{\rm ML} = 1.5$ in the Solar case ($Z=0.02$ in their models). The comparison with observations shows that the position of the blue IS boundary in these models is generally too cold, over the entire luminosity range at Solar metallicity and at low luminosities in the case of the MCs. However, the red edges at LMC and SMC metallicity agree well with our predictions. Last, but not least, we overlay IS boundaries computed using MESA from \citet{Deka2024} who tested several assumptions related to convection. Figure \ref{fig:IS} shows their set A, which is a simple convection model, and their set D, where all radiative cooling, turbulent pressure, and turbulent flux are included simultaneously. Unfortunately, the \teff-$L$ relations provided in \citet{Deka2024} represent envelopes to models spanning the metallicity range 
$Z \in \{0.013, 0.006, 0.002\}$ (or \mh $\in \{0.00, -0.34, -0.75\}$) and thus cannot be assessed for metallicity effects. Set D in particular is much wider than allowed by the empirical constraints in the MW and the LMC. The narrower set A matches the locations of our predicted IS at SMC metallicity and otherwise predicts a bluer boundary than supported by observations.

In our approach, all models in the three-dimensional grid ($\mathcal{M}$, $Z$, $\omega$) are initialized at the ZAMS and evolutionary tracks are calculated until the end of core Helium burning. This ensures consistency of the computed models because physical effects that modify the mass-luminosity relation, such as overshooting, rotation, and mass loss, are considered throughout the evolution of the star. Conversely, ad hoc increases in luminosity of Cepheid models as implemented in cases B \& C in \citet{DeSomma2020,DeSomma2022} cannot be attributed to specific physical origins and do not self-consistently track the influence of convective core overshooting, for example, on the evolution of the star. 

\subsection{Computing synthetic Cepheid populations with \texttt{SYCLIST}}
\label{sec:syclist}

We computed synthetic populations using the Geneva population synthesis code \texttt{SYCLIST} \citep{Georgy2014}, which we adapted to output data required to distinguish the properties of primary and secondary stars in binary systems. The simulated populations mimic a constant star formation rate by adding together 296 synthetic star clusters with a mass of $50\, \cdot 10^3 \, \msun$ in the age range [5, 300] Myr, i.e., using a time step of 1 Myr, for each of the three metallicity values $Z=0.014$, 0.006, and 0.002.
We used \texttt{SYCLIST} default parameters, such as a \citet{Salpeter1955} initial mass function between 1.7 and 15 \msun, the angular velocity distribution from \citet{Huang2010}, a random spatial distribution of the rotation axis of the star $\sin (i)$ for the angle of view, a gravity darkening law following \citet{EspinosaLara2011}, limb darkening based on \citet{Claret2000}, and a $30\%$ probability of a star being in a binary system. 

\begin{figure*}
    \centering
    \includegraphics[width=\hsize]{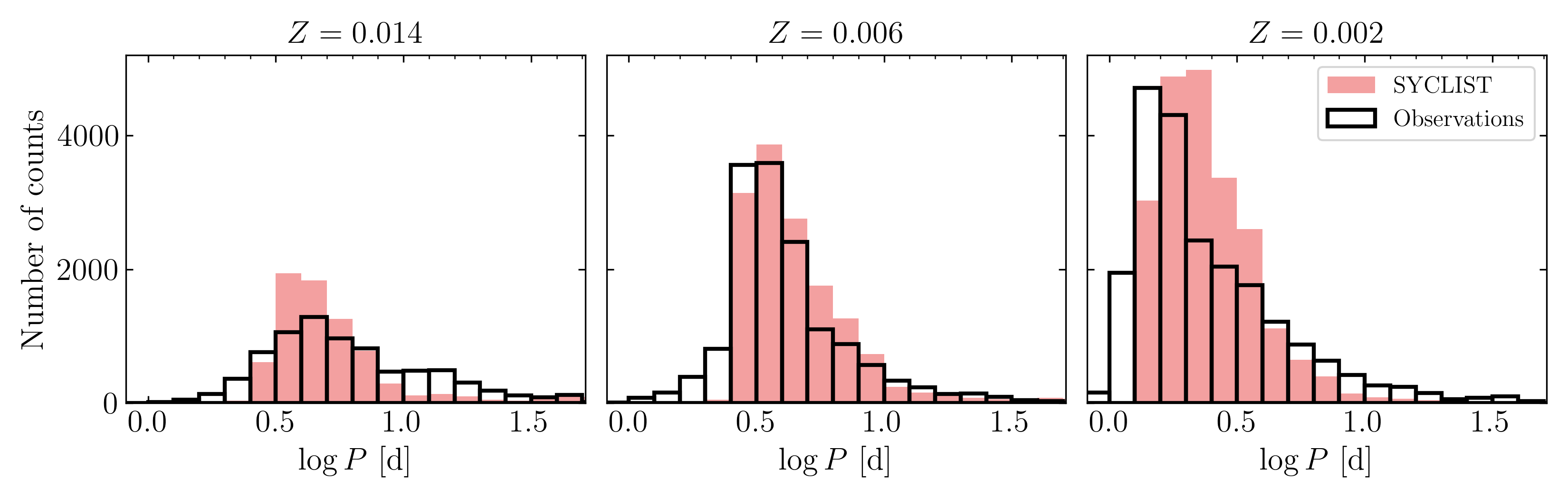}
	\caption{Distribution of $\log P$ for $Z=0.014$ (left), 0.006 (middle), and 0.002 (right panel). The \texttt{SYCLIST} populations appear in red, while observations are shown as black histograms. Observations come from \citet{Pietrukowicz2021} and \citet{Soszynski2015,Soszynski2017} for Classical Cepheids in the Milky Way and Magellanic Clouds, respectively. Observational counts have been rescaled in order to match the numbers we have in \texttt{SYCLIST} simulations, by a factor 3.4, 5.9, and 7.5 for $Z=0.014$, 0.006, and 0.002, respectively. The age limit applied ($< 300$\,Myr) to the synthetic populations leads to mismatches at the shortest periods, particularly at $Z=0.002$.}
	\label{fig:logP_distribution}
\end{figure*}

\texttt{SYCLIST} relies on single star evolution tracks computed with the Geneva stellar evolution code. Hence, the treatment of binary stars is simplistic, and neither binary evolution \citep{Neilson2015,Karczmarek2023,Dinnbier2024} nor cluster dynamics \citep{Dinnbier2022} are taken into account. Multiplicity is tagged by the \texttt{Bin} flag, which can take four different values. Single stars are flagged as \texttt{Bin==0}. Stars flagged with \texttt{Bin==2} are cases where the mass of the secondary is lower than the minimum mass allowed by the grid of models, and we decided to remove them from our sample. \texttt{Bin==1} and \texttt{Bin==3} correspond to the primary and secondary components of a binary system.

We considered as Cepheid any (individual) non-MS ($\log{g} < 3.4$) star whose luminosity $L$ and effective temperature \teff\ fall within the IS boundaries determined using the same models \citepalias{Anderson2016}.
For binary systems, we consider $L$ and \teff\ of individual components to determine whether a star is a Cepheid. 
The blue edge is averaged over the second and third crossings, while for the red edge we use both boundaries corresponding to first crossing and second/third crossings and the transition occurs where they intersect with one another: at ($\log \teff$, $\log L$) values of (3.72, 3.19), (3.74, 3.14), and (3.76, 2.76) for the MW, LMC, and SMC, respectively. If needed, IS edges are extended to lower/higher luminosities in order to include stars falling outside of their range. The position of blue loops is imprinted on the location of stars in these HRDs, as can be seen from the red dashed lines in Fig. \ref{fig:SYCLIST_HRD}. We limit the number of first crossing Cepheids by only considering stars more luminous than these lines, defined by the following two ($\log \teff$, $\log L$) points \{(3.7962, 3.027), (3.7510, 2.7662)\}, \{(3.8035, 3.027), (3.7564, 2.7698)\}, \{(3.8194, 2.610), (3.7761, 2.4952)\} for $Z=0.014$, 0.006, and 0.002, respectively. These cuts are defined where there is an apparent drop in the density of stars as we move towards a lower luminosity level along the IS, which we associate to the extremely rapid Hertzsprung gap phase. We note that this distinction becomes increasingly harder as one moves to lower metallicities, for instance at $Z=0.002$, because the minimum mass of Cepheids on second/third crossings decreases with decreasing metallicity \citepalias{Anderson2016}. Hence, some contamination might remain in our selection but should be fairly negligible. We have a total of 8332, 15043, and 24289 Cepheids selected for $Z=0.014$, 0.006, and 0.002, respectively. 
The resulting Cepheid populations, whose HRDs are illustrated on Fig. \ref{fig:SYCLIST_HRD}, are well sampled and thus suitable for determining the effect of metallicity on LLs and can be compared to key features of observed populations. To this end, we generally compare MW Cepheids to predictions at Solar metallicity ($Z=0.014$), LMC Cepheids to $Z=0.006$, and SMC Cepheids to $Z=0.002$. However, we caution that observed Cepheid populations are generally more complex than our synthetic populations, notably with respect to star formation history, non-unique values of metallicity, and stellar multiplicity. Additionally, we caution that many MW Cepheids with high-quality spectroscopic observations have super-Solar abundances, exceeding the range of the models.

\begin{figure*}
    \centering
    \includegraphics[width=\hsize]{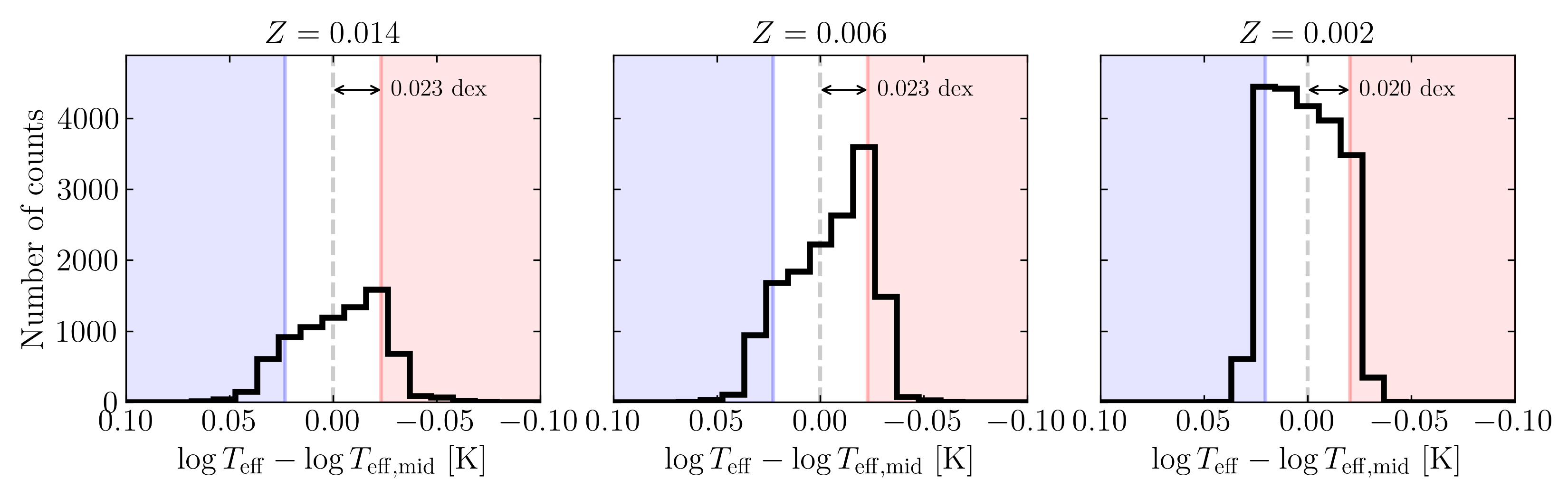}
	\caption{Distribution of $\log \teff-\log T_{\rm eff, mid}$ for our selection of IS stars in \texttt{SYCLIST} simulations, where $\log T_{\rm eff, mid}$ corresponds to the effective temperature at the mid-IS line right in the middle between the blue and red edges of the IS, for $Z=0.014$ (left), 0.006 (middle), and 0.002 (right panel). The vertical lines indicate the $1\sigma$ dispersion that would correspond to a uniform distribution of stars within the IS.}
	\label{fig:IS_scatter}
\end{figure*}

Luminosities were translated to magnitudes in several photometric passbands to enable the comparison with observations. 
We use empirical calibrations by \citet{Worthey2011} to compute colours and magnitudes. The latter take \logg, \mh, and \teff as input parameters, with $M_{\rm bol, \odot}=4.75$ mag, and return several colour indices, i.e. $U-B$, $B-V$, $V-R$, $V-I$, $J-K$, $H-K$, and $V-K$, as well as the bolometric correction for the $V$-band magnitude. Using predicted combinations of $V$-band magnitudes and colours, we obtain magnitudes for the $I$, $J$, $H$, and $K_s$ filters. Colour transformations based on $V-I$\footnote{Table 5.7 at \url{https://gea.esac.esa.int/archive/documentation/GEDR3/Data_processing/chap_cu5pho/cu5pho_sec_photSystem/cu5pho_ssec_photRelations.html}} allow us to derive \gaia magnitudes $G$, $G_{\rm BP}$, and $G_{\rm RP}$. We also make use of the transformations between ground-based and \textit{Hubble} Space Telescope (\textit{HST}) system apparent magnitudes empirically determined by \citet[][Sect. 2.3]{Breuval2020}, in the F160W, F555W, and F814W passbands. Furthermore, we used the \textit{JWST} magnitude conversion tool\footnote{\url{https://github.com/spacetelescope/jwst_magnitude_conversion}} to compute magnitudes in the NIRCam F090W, F115W, F150W, F277W, F356W, and F444W bands, using predicted $V$ and $K_s$-band magnitudes. 
Using combinations of various passbands, we compute the following reddening-free Wesenheit magnitudes \citep{Madore1982} assuming a \citet{Fitzpatrick1999} reddening law with $R_V = 3.1 \pm 0.1$, following \citetalias[][see their Table 3]{Breuval2022}:
\begin{align}
    W_{\rm G} &= G - 1.900 \: (G_{\rm BP}-G_{\rm RP}) \\
    W_{\rm VI} &= I - 1.387 \: (V-I) \\
    W_{\rm H} &= \rm{F160W} - 0.386 \: (\rm{F555W}-\rm{F814W}) \label{eq:SH0ES_mWH}\\
    W_{\rm VK} &= K_s - 0.127 \: (V-K_s) \\
    W_{\rm JK} &=  K_s - 0.735 \: (J-K_s) \ .
\end{align}
In the case of non-spherical stars, \texttt{SYCLIST} applies corrections to predicted values of $L$ and \teff to account for the effects of: the angle of view, the angular velocity, gravity darkening, and limb darkening. Effects on these stellar properties are at the level of $\sim 0.1$\% for the luminosity, and $\sim 0.01$\% for the effective temperature for Cepheids. These corrections are discussed in more detail in Sects. 2.3.1 to 2.3.3 of \citet{Georgy2014}. 
We discarded simulated Cepheids in binary systems where the constant flux of a companion star would significantly bias the Cepheid's amplitude by more than 0.2 mag. 
Strongly contaminated Cepheids would typically not be discovered due to insufficient variability or cut from observed samples via $\sigma$ clipping \citep[cf.][]{Anderson2018}.

We computed fundamental-mode pulsation periods, $P$, following the tight period-radius relations consistently derived by \citetalias[][]{Anderson2016} (see their Table 5) using the identical models. 
Averaging over different rotation rates,
the blue and red edges of the IS, and the second and third crossings, we obtained metallicity-dependent fiducial period-radius relations at the center of the IS:
\begin{align}
    \log P = (\log R - 1.166) / 0.668 \; \textrm{for $Z=0.014$,} \label{eq:PR_1} \\
    \log P = (\log R - 1.142) / 0.675 \; \textrm{for $Z=0.006$,} \label{eq:PR_2} \\
    \log P = (\log R - 1.122) / 0.692 \; \textrm{for $Z=0.002$,}
    \label{eq:PR_3}
\end{align}
with $1\sigma$ dispersions of 0.025, 0.032, and 0.033, respectively. We accounted for the intrinsic width of the period-radius relation by adding a random offset to each period computed using Eqs.\,(\ref{eq:PR_1}-\ref{eq:PR_3}). The added offsets were normally distributed with zero mean and a standard deviation corresponding to one quarter of the mean period difference between the red and blue edges in the $\log{R}$ interval $[{1.8, 2.1}]$ for the same metallicity value: 0.0148, 0.0217, and 0.0195 for $Z=0.014$, 0.006, and 0.002, respectively. This is equivalent to assuming that the blue and red IS edges encompass $>99\%$ of all possible periods.

\section{Properties of the synthetic Cepheid populations}
\label{sec:properties}

\subsection{Predicted vs observed period distributions}\label{sec:pdist}
Figure\,\ref{fig:logP_distribution} compares our Cepheid populations to the observed fundamental mode Cepheid period distributions of the MW \citep{Pietrukowicz2021}, LMC \citep{Soszynski2015}, and SMC \citep{Soszynski2017}. We generally found good agreement between observed and predicted period distributions considering the caveats involved in the comparison, notably concerning metallicity differences, with the peak of the distribution moving to shorter periods as the metallicity decreases. The modes of the predicted distributions agree well for the MW and the LMC. This is particularly important because the peak is dominated by the minimum mass at which models exhibit blue loops that enter the IS. 
The predicted distribution for $Z=0.002$ suggests a slowly increasing number of Cepheids at periods longer than the observed distribution's mode. This artefact of the age limit ($< 300$\,Myr) applied here \citepalias[e.g., a $2.5$\,d $Z=0.002$ Cepheid at the IS center, crossing averaged, is roughly $350$\,Myr old, cf.][their Table 4]{Anderson2016} has no bearing on the distance scale.

At periods longer than the peak, both simulated and observed distributions decline exponentially, and there is good agreement for the two lower metallicity cases. The broader distribution of MW periods likely reflects the larger metallicity range of its constituent Cepheids and the more complex star formation history. 
As a sidenote, \citet{Anderson2017} previously showed a more significant mismatch in period distributions for first overtone Cepheids compared to fundamental mode Cepheids, although the origin of this mismatch remains unknown.

\subsection{How Cepheids sample the instability strip}\label{sec:ISscatter}

Evolutionary timescales vary along the blue loop. This could lead to a non-uniform sampling of the instability strip. Here, we compare the distributions of Cepheids in our synthetic population across the IS to a uniform distribution based on the positions of the IS edges.

\begin{figure*}
    \centering
    \includegraphics[width=\hsize]{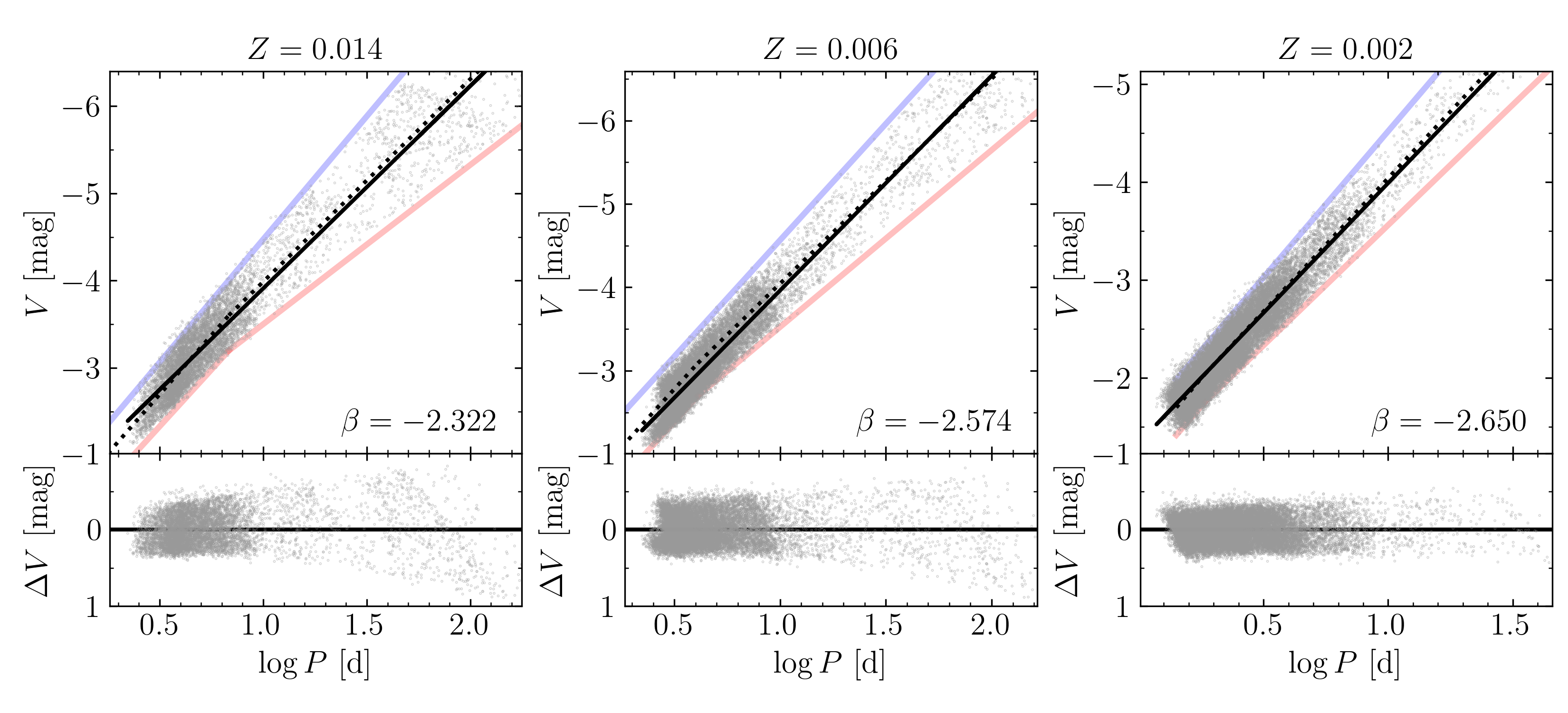}
	\caption{\textit{Top:} LLs in the $V$ band, for $Z=0.014$ (left), 0.006 (middle), and 0.002 (right panel). IS stars from \texttt{SYCLIST} are shown as grey points. The blue and red lines correspond to the IS boundaries from \citetalias{Anderson2016}. The black solid line is the LL fit to the \texttt{SYCLIST} population, for $\log P_{\rm 0} = 0.0$. The corresponding slope value, $\beta$, is annotated on the bottom right of each panel. The black dashed line shows the mid-IS fit, obtained by averaging the magnitude values for the blue and red boundaries at a given period. It shows a break in slope where the transition between the red edges for first crossing and second+third crossings occurs. Note the different axis ranges. \textit{Bottom:} Residuals computed as the difference in magnitude between the individual \texttt{SYCLIST} stars and that of the \texttt{SYCLIST} LL fit at the same period.}
	\label{fig:PLRs_V}
\end{figure*}

We define a mid-IS line that bisects the blue and red edges, with a break at the intersection between the first crossing and the second/third crossings. Following this, we consider two different distributions:

\begin{figure*}
    \centering
    \includegraphics[width=\hsize]{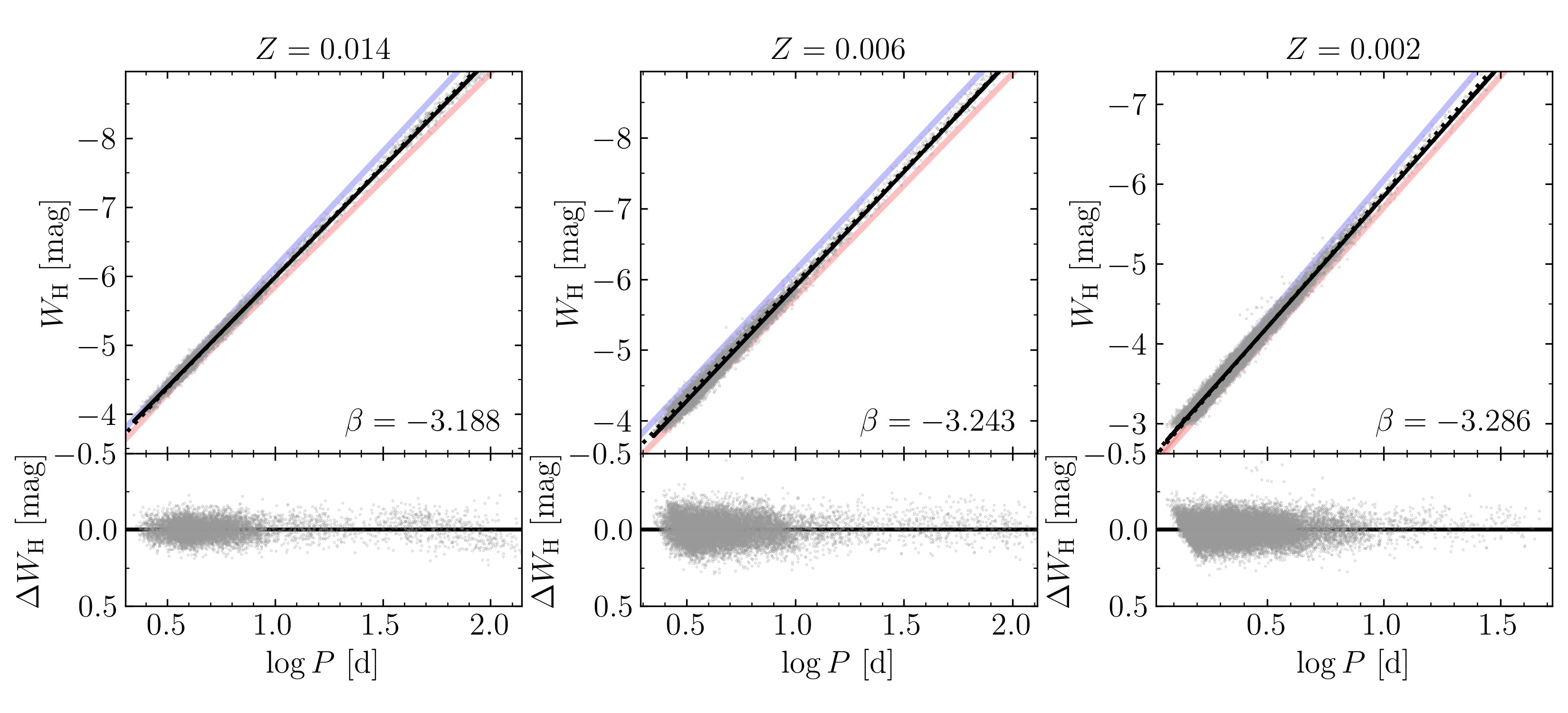}
	\caption{Same as Fig. \ref{fig:PLRs_V} for the $W_{\rm H}$ magnitude}
	\label{fig:PLRs_WH}
\end{figure*}

\begin{itemize}
    \item a uniform distribution between the blue and red edges of the IS. For this, we uniformly generated 10,000 $\log L$ values between the minimum and maximum, i.e. [1.1969, 4.9071], [1.362, 4.9969], and [1.4983, 5.0657] for $Z=0.014$, 0.006, and 0.002, respectively. We then compute the corresponding $\log \teff$ on the blue and red sides, either based on the first or second/third crossings depending on the luminosity, and again generate temperatures uniformly distributed between those edges;
    \item the distribution obtained from \texttt{SYCLIST}.
\end{itemize}
We computed the difference between the $\log \teff$ of the above two distributions and that of the mid-IS line, which shows a break where the red boundary of the IS changes slope, so that the resulting distributions are centred on zero. Then, a positive value would mean that the star is located more towards the blue edge of the IS (hotter \teff), while a negative value suggests that it falls on the redder side of the IS (cooler \teff). This is illustrated in Fig. \ref{fig:IS_scatter}. We would expect to see a roughly flat distribution if the IS was uniformly populated. There does not seem to be a significant predominance of bluer or redder stars for $Z=0.002$. However, for $Z=0.006$, and to a lesser extent $Z=0.014$, stars seem to pile up towards the red IS edge. Hence, our \texttt{SYCLIST} populations demonstrate that Cepheids are not uniformly distributed in \teff within the IS.

\begin{figure*}
    \centering
    \includegraphics[width=\hsize]{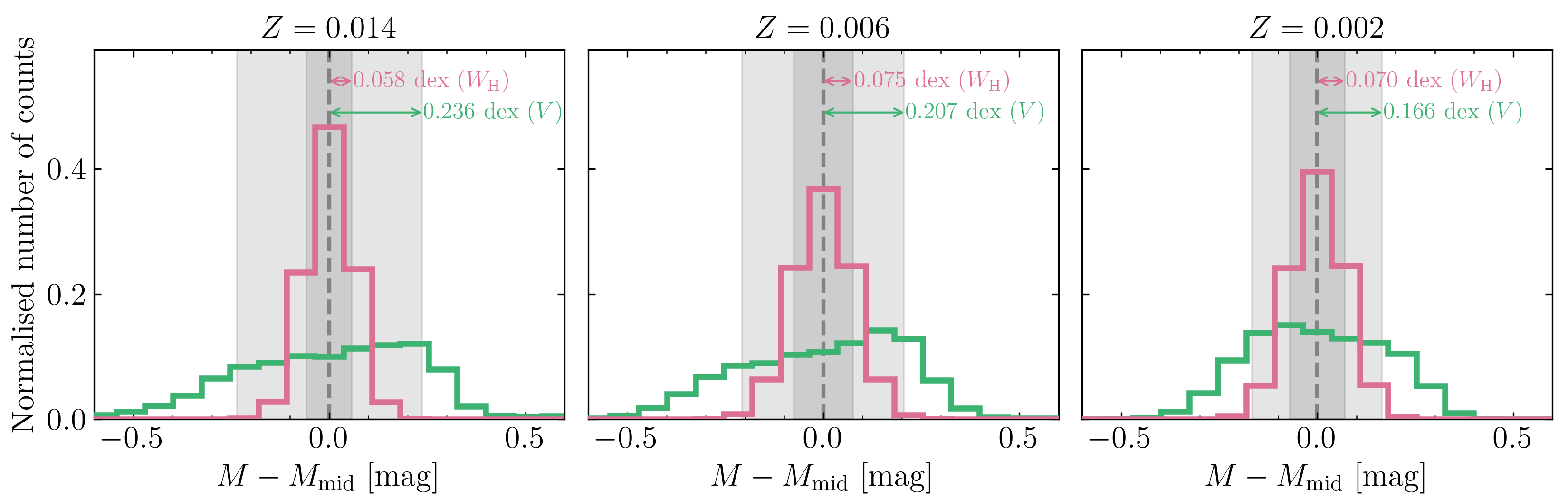}
	\caption{LL scatter in the $V$-band (green) and $W_{\rm H}$ magnitudes (pink), for $Z=0.014$ (left), 0.006 (middle), and 0.002 (right panel). The histograms show the difference in magnitude between the individual \texttt{SYCLIST} IS stars and that of the \texttt{SYCLIST} LL fit at the same $\log P$. The shaded regions correspond to the standard deviation, hence to the scatter of the LL and the values are reported on the plot with the corresponding colour.}
	\label{fig:PLRs_scatter_V_WH}
\end{figure*}

\subsection{The intrinsic scatter of the Leavitt law \label{sec:ISwidth}}

We proceed with the translation of modelled quantities, i.e. the intrinsic luminosity and radius of the star, to observables that can then allow us to make comparisons with observational studies. We fit linear LLs for each combination of photometric band and metallicity to the \texttt{SYCLIST} populations of IS stars. After which, we fit the resulting intercept and slope parameters to obtain $\alpha_{\mathrm{M}}$ and $\beta_{\mathrm{M}}$. To this end, we use the formalism from Sect. \ref{sec:intro} (Eqs. \ref{eq:pl_general}-\ref{eq:plz_full}). We also cross-check these results against the mid-points of the individual IS edges, where LL are first fitted separately on the blue and red edges and blue-red averaged intercept and slope are then used to derive metallicity parameters.

\begin{table}[]
    \caption{Width of the instability strip, as measured in the observations and predicted with our models.}
    \label{table:WIS_SYCLIST}
    \centering
    \small
    \begin{tabular}{@{}c | c c | c c c@{}}
        \hline\hline
        Band         & Obs.  & Notes                        & $Z=0.014$  & $Z=0.006$  & $Z=0.002$   \\
                     & [mag]         &                      & [mag]      & [mag]      & [mag]       \\
        \hline
        $G_{\rm BP}$ & 0.23          & A/Ri19                & 0.246      & 0.218      & 0.176       \\
        $V$          & 0.22          & B/M06                & 0.236      & 0.207      & 0.166       \\
        $G$          & 0.19          & A/Ri19                & 0.208      & 0.188      & 0.153       \\
        $G_{\rm RP}$ & 0.16          & A/Ri19                & 0.171      & 0.155      & 0.125       \\
        $I$          & 0.14          & B/M06                & 0.165      & 0.151      & 0.122       \\
        F090W        & -             &  -                   & 0.146      & 0.133      & 0.111       \\
        F115W        & -             &  -                   & 0.118      & 0.113      & 0.096       \\
        $J$          & 0.11          & A/P04                & 0.112      & 0.110      & 0.095       \\
        F150W        & -             &  -                   & 0.087      & 0.092      & 0.082       \\
        $H$          & 0.09          & A/P04                & 0.073      & 0.084      & 0.076       \\
        $K_s$        & 0.07          & A/P04                & 0.067      & 0.080      & 0.074       \\
        F277W        & -             &  -                   & 0.070      & 0.081      & 0.074       \\
        F356W        & -             &  -                   & 0.065      & 0.079      & 0.073       \\
        F444W        & -             &  -                   & 0.071      & 0.082      & 0.075       \\
        \hline
        $W_{\rm G}$  & 0.10          & A/Ri19                & 0.077      & 0.086      & 0.075       \\
        $W_{\rm VI}$ & 0.077         & A/S15                & 0.079      & 0.087      & 0.075       \\
        $W_{\rm H}$  & 0.069         & A/SH0ES              & 0.058      & 0.075      & 0.070       \\
        $W_{\rm VK}$ & 0.077         & A/Ri22                & 0.054      & 0.074      & 0.069       \\
        $W_{\rm JK}$ & 0.086         & A/Ri22                & 0.051      & 0.073      & 0.068       \\
        \hline
    \end{tabular}
    \tablefoot{There are only small differences, at the 0.005 mag level, in the dispersion predicted when one accounts for the flux contribution from companion stars or only for that of the primary stars alone so we decided to only show the former set of values here. A: LMC, B: NGC 4258; Ri19: \citet{Ripepi2019}, M06: \citet{Macri2006}, P04: \citet{Persson2004}, S15: \citet{Soszynski2015}, SH0ES: \citet{Riess2019}, Ri22: \citet{Ripepi2022}.}
\end{table}

Figures \ref{fig:PLRs_V} and \ref{fig:PLRs_WH} show the LL fits obtained for the $V$-band and $W_{\rm H}$ magnitudes. We highlight the fits based on the \texttt{SYCLIST} population of IS stars, but also show those corresponding to the blue and red edges separately, as well as the average between the two with the break where the red slope switches from first crossing to second/third crossings. Slight differences appear between the \texttt{SYCLIST} and mid-IS fit, and their impact on the resulting metallicity parameters $\alpha_{\mathrm{M}}$ and $\beta_{\mathrm{M}}$ are discussed in the following section. These differences are to be expected because the IS is not uniformly populated across temperature as shown in Sect.\,\ref{sec:ISscatter}. 

For the SH0ES Wesenheit magnitude, $W_{\rm H}$, we find a slope of $-3.243$ at LMC metallicity and note that this is in very close agreement with the global LL slope adopted in \citetalias[$-3.299 \pm 0.015 \, \rm mag \, dex^{-1}$]{Riess2022}, which is strongly constrained by 339 LMC Cepheids with precise photometry from the ground and from {\it HST}. Figures of LL fits for other magnitudes are provided via Zenodo\footnote{\url{https://doi.org/10.5281/zenodo.15495081}}, and the fits parameters are provided for the \texttt{SYCLIST} populations either using a free or fixed slope, the blue and red IS edges separately, and the midline between the two IS edges in App. \ref{app:PLRs}: Tables \ref{table:PLR_SYCLIST}, \ref{table:PLR_SYCLIST_fixed}, \ref{table:PLR_ISedges_blue}, \ref{table:PLR_ISedges_red}, and \ref{table:PLR_ISedges_mid}, respectively.

\begin{figure*}
    \centering
    \includegraphics[width=\hsize]{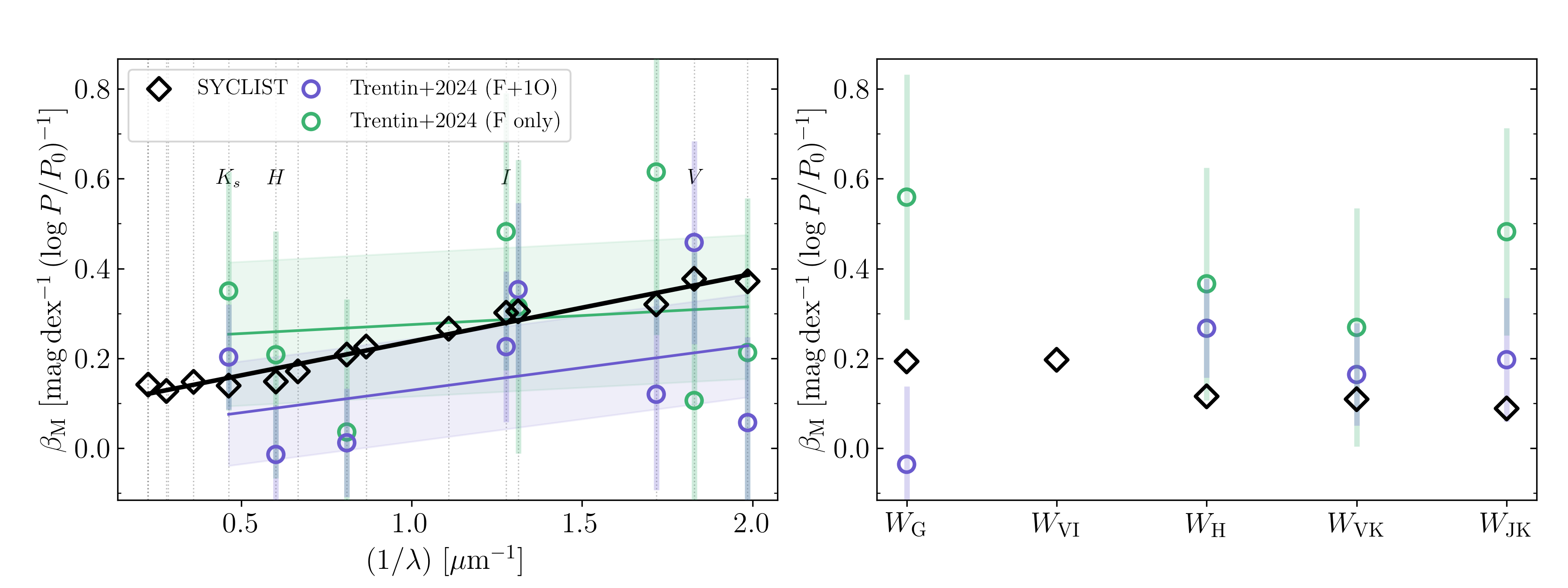}
	\caption{Metallicity effect on the LL slope, $\beta_{\mathrm{M}}$, as defined in Eq.\,(\ref{eq:plz_full}). The left panel shows $\beta_{\rm M}$ as a function of the effective wavelength $1/\lambda$, while the right panel shows the comparison for Wesenheit magnitudes. Our predictions based on \texttt{SYCLIST} are shown in black. Observational results from \citet{Trentin2024} are shown for the fundamental mode and first overtone Cepheids and fundamental mode Cepheids alone in blue and green, respectively. In the left panel, the vertical dotted lines show the location of all the photometric passbands considered, and we label some of them to orient the reader.}
	\label{fig:LL_metallicity_slope}
\end{figure*}

We consider the LL scatter as an important diagnostic for the quality of our synthetic populations. It is an empirical fact that arises from the width of the IS. Observations exhibit a decrease in scatter as one moves towards redder filters and also for Wesenheit magnitudes  \citep[e.g.,][]{Madore2012}. This wavelength dependence can be attributed to the connection between the intrinsic LL scatter and the period-luminosity-colour relation. As expected, the LL scatter for the $W_{\rm H}$ magnitude is lower with respect to the one for the $V$ band. The overall dispersion gets slightly smaller as the metallicity decreases when using the $V$ band, but we have the opposite effect for $W_{\rm H}$. Figure \ref{fig:PLRs_scatter_V_WH} shows the distribution of the residuals, as an illustration of the LL scatter, for the $V$ and $W_{\rm H}$ magnitudes. We find $0.207$ and $0.075$ mag for the $V$ and $W_{\rm H}$ magnitudes at LMC metallicity, while observations have reported values of $0.22$ and $0.069$ mag. The latter value was used as a finite width for the $W_{\rm H}$ LL, and obtained by subtracting quadratically the photometric measurement errors (0.03 mag) to the observed dispersion \citepalias[0.075 mag,][]{Riess2022}. We provide in Table \ref{table:WIS_SYCLIST} the predicted scatter values we obtain for all the magnitudes, at each metallicity value, and also quote the observationally-derived scatters for comparison. We find that there is a generally very good agreement between these two. Our predicted scatters fall typically within $\pm 0.01 - 0.02$\,mag of the observed values, when comparing with $Z=0.006$, given that most studies focused on the LMC. Furthermore, we find the same trend whereby the scatters are larger for bluer wavelengths, and get smaller as we go towards redder filters. Also, even if we were to consider single stars only in our simulations, by not accounting for the flux contribution of companion stars, the impact on our predicted scatters would be negligible.

\section{Metallicity dependence of the Cepheid LL}
\label{sec:LL_metallicity}

The previous sections established that the models considered here provide a very close match to a)  the position of the IS, b) the period distribution of Cepheids at the respective metallicities, and c) the width of the IS across many wavelengths. Additionally, the slope of the predicted $W_{\rm H}$ LL for LMC metallicity is a close match to the global slope adopted in SH0ES \citepalias{Riess2022}. Furthermore, \citetalias{Anderson2016} demonstrated very good agreement among predicted and observed a) period changes \citep[recently also shown using period changes measured from radial velocities;][]{Anderson2024}, b) masses, c) radii, and d) flux-weighted gravity \citep{Anderson2020corr,2020A&A...640A.113G}. The following compares the predicted dependence of the LL on metallicity to observational results reported in the recent literature. Figures \ref{fig:PLRs_V}, \ref{fig:PLRs_WH} illustrate LLs derived from our synthetic populations; additional figures are available via Zenodo.

\begin{figure}
    \centering
    \includegraphics[width=\hsize]{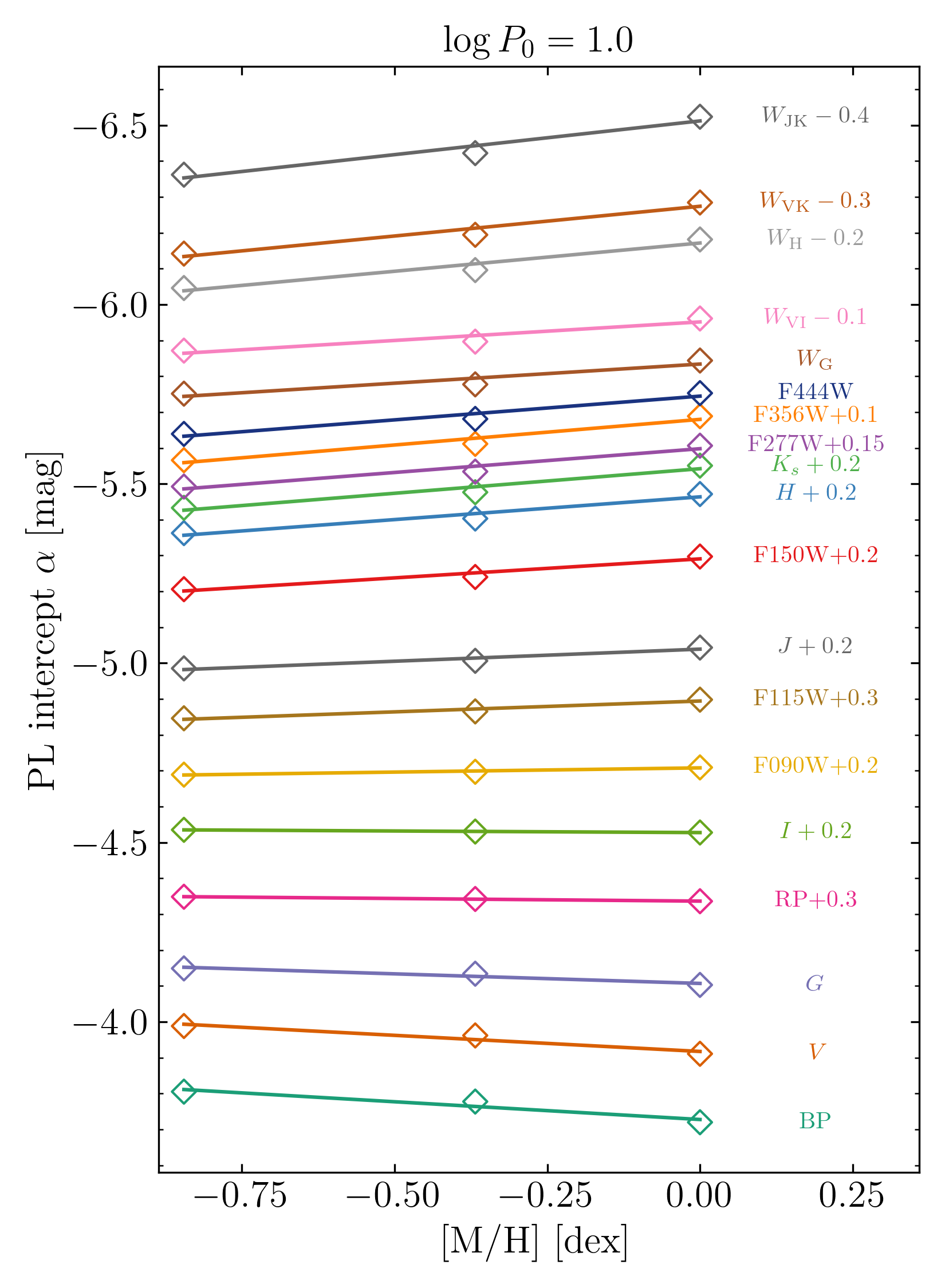}
	\caption{LL intercepts $\alpha$ determined from synthetic Cepheid populations as a function of metallicity \mh for the different single-wavelength and Wesenheit magnitudes considered, and for $\log P_0 = 1.0$. Some lines are shifted vertically, as indicated by the labels, in order to improve the readability of the plot. See Table \ref{table:plz_syclist} for the $\alpha_{\rm M}$ values associated to these fits.}
	\label{fig:PLRs_alpha}
\end{figure}

\subsection{The LL slope-metallicity dependence\label{sec:LL_slope_metallicity}}
A predicted slope-metallicity dependence ($\beta_M \ne 0$) was first reported by \citetalias{Anderson2016} and confirmed by other stellar models \citep{DeSomma2022}. Here, we find that our synthetic LLs also exhibit a slope-metallicity dependence in all individual filters and Wesenheit magnitudes, in the sense that LL slopes become steeper (more negative) at lower metallicity, i.e., $\beta_{\mathrm{M}} > 0$ in Eq.\,(\ref{eq:plz_full}). 
We fit the slope component of Eq. (\ref{eq:plz_full}) to derive $\beta_{\mathrm{M}}$ for individual filters and Wesenheit magnitudes, as reported in Table \ref{table:plz_syclist}. The left panel of Fig.\,\ref{fig:LL_metallicity_slope} illustrates our \texttt{SYCLIST} predictions as a function of wavelength. They range from $\sim 0.13$, at redder wavelengths, to $\sim 0.38$, at bluer wavelengths. 

The predicted slope-metallicity dependence is supported by observations. Table~2 in \citet[OGLE-III Magellanic Cloud Cepheids]{Soszynski2015} reveals that the LL slope is steeper among the lower-metallicity SMC Cepheids in all of $I-$, $V-$, and Wesenheit $W_{VI}$ magnitudes for Cepheids pulsating in the fundamental mode, first, and second overtone. Although the difference reaches $> 10\sigma$ significance in the optimal case of fundamental mode Cepheids in $W_{VI}$, no note on this issue was made. Table~4 in \citet{Breuval2021} lists LL slopes across a wider range of passbands for fundamental mode Cepheids. While the reported uncertainties precluded establishing statistically significant trends, we noticed that all Wesenheit LLs yielded steeper slopes in the SMC compared to the LMC. Metallicity-dependent slopes have also been investigated using MW Cepheids with direct spectroscopic abundance measurements, although no significant effect has been established for now \citep[e.g.,][]{Ripepi2020,Trentin2024}. 

Nonetheless, it is worth pointing out that Figs.~4 and A1 in \citet{Trentin2024} suggest a preference for a slope dependence (their parameter $\delta$) on the order of $\beta_{\mathrm{Fe}} \approx 0.1$\,mag\,dex$^{-1}\,(\log{P/P_0})^{-1}$ for fundamental mode and first overtone Cepheids and $\beta_{\mathrm{Fe}} \approx 0.2$\,mag\,dex$^{-1}\,(\log{P/P_0})^{-1}$ for fundamental mode Cepheids alone. Their $\beta_{\mathrm{Fe}}$ are also displayed on Fig. \ref{fig:LL_metallicity_slope}. Although their approach differs from ours, in that they only use MW Cepheids with a broad range of \feh, their results still show a very good agreement with our model predictions. We note that \citet{Trentin2024} reported a $\sim 0.1$\,mag\,dex$^{-1}$ lower value for $\alpha_{\mathrm{Fe}}$ when neglecting the slope-metallicity dependence. 

\subsection{The LL intercept-metallicity dependence\label{sec:LL_intercept_metallicity}}
As mentioned in Sect.\,\ref{sec:LL_slope_metallicity}, $\alpha_\mathrm{M}$ depends on pivot period if $\beta_{\mathrm{M}}\ne 0$. Values of $\alpha_\mathrm{M}$ must therefore be accompanied by the corresponding $P_0$. Additionally, care must be taken to compare predictions to observational studies based on equivalent approaches, since the majority of observational studies assume a universal LL slope (i.e., $\beta_{\mathrm{M}}=0$). To facilitate comparisons with observational studies, we report our results in Table\,\ref{table:plz_syclist} for $\log{P_0}=0.0$, $0.7$, and $1.0$ for a wide range of passbands from optical to infrared wavelengths. 

Figure \ref{fig:PLRs_alpha} illustrates the metallicity dependence of the LL intercept $\alpha$, for $\log P_0 = 1.0$, with $\beta$ a free parameter. $\alpha$ exhibits a mild dependence on metallicity, which increases with wavelength and is nearly flat around the $I-$band. The predicted trends with \mh appear to be non-linear, in particular at the shortest and longest wavelengths, and  
flatten off towards lower \mh, resembling the trend reported based on observations in \citet[their Fig.~8]{Breuval2024}. Nevertheless, we determined $\alpha_{\mathrm{M}}$ using a linear fit to the three points, in keeping with the literature. Predictions at super-Solar \mh would usefully constrain a possible non-linear trend.

Figure\,\ref{fig:lambda_gamma} illustrates the wavelength dependence of the LL intercept-metallicity effect $\alpha_{\mathrm{M}}$ determined from the linear fits in Fig.\,\ref{fig:PLRs_alpha}.  
The left panel shows our predictions for $\log P_{\rm 0} = 0.7$ to enable a direct comparison with the observational study by \citetalias{Breuval2022}. Predictions based on \texttt{SYCLIST} populations are shown in black when $\beta$ is a free parameter and in pink when $\beta$ is fixed by our $Z=0.006$ (LMC) results. Predictions based on the IS midline are shown in grey. \texttt{SYCLIST} populations yield a slightly steeper wavelength dependence of $\alpha_{\mathrm{M}}$, although the difference is hardly significant, reaching a maximum of $0.03$\,mag\,dex$^{-1}$ at the shortest wavelengths. Fixing the slope does not affect the predicted wavelength dependence significantly. Table \ref{table:plz_syclist} tabulates the predictions based on the \texttt{SYCLIST} populations shown in Fig.\,\ref{fig:lambda_gamma}, Tab. \ref{table:gamma_IS} the values for predictions based on the IS bisector.

The observational study by \citetalias{Breuval2022} relied on the metallicity differences between MW, LMC, and SMC Cepheids to investigate $\alpha_\mathrm{Fe}$ effects using an LL slope fixed by the Cepheids in the LMC and $\log{P_0}=0.7$. We considered this study to be the most similar in approach to our predictions, since each of the three Cepheid groups featured a small range in metallicity, and the determination of absolute magnitudes was particularly robust with respect to distance precision and (comparatively) low reddening. Further discussion of the varying level of (dis-)agreement among empirical studies has been provided by \citetalias{Breuval2022}, \citet{Breuval2024}, \citet{Bhardwaj2024}, and \citet{Trentin2024}, for example. Comparing measurements from \citetalias{Breuval2022} to our \texttt{SYCLIST} predictions with fixed LL slope (pink) reveals some disagreements at the level of $0.1-0.2$\,mag\,dex$^{-1}$, with our values varying with $1/\lambda$ from $-0.19$ to $-0.02$ \,mag\,dex$^{-1}$, while the values in \citetalias{Breuval2022} are more negative and exhibit a weaker dependence on $1/\lambda$. At $1\mu$m, the difference is approximately $0.15$\,mag\,dex$^{-1}$. We note that values attributed to \citetalias{Anderson2016} as reported in \citetalias{Breuval2022} were determined for $\log{P_0}=0.0$, rendering them more negative. Furthermore, we note that Sect. 5.4 in \citetalias{Breuval2022} shows that a more negative SMC metallicity ([Fe/H] $=-0.90$ instead of $-0.75$) would tend to increase $\alpha_{\mathrm{Fe}}$ by $\sim 0.05$\,mag\,dex$^{-1}$, resulting in better agreement with our predictions. However, the SMC's updated iron abundance of $-0.785 \pm 0.012$ ($\sigma=0.082$) reported in \citet{Breuval2024} with attribution to Romaniello et al. 2024 (in prep.) does not support this. Instead, the order of magnitude and wavelength dependence of the difference with \citetalias{Breuval2022} could suggest a link to reddening. For example, the $0.2$\,mag\,dex$^{-1}$ difference in $V-$band corresponds to a $0.15$\,mag difference assuming a $0.75$\,dex metallicity lever. Given the significant reddening of MW Cepheids \citep[mean $E(B-V)=0.5$ in][]{Breuval2022}, this difference could be explained by even a modest uncertainty in either $E(B-V)$ or $R_V$. Further work is required to investigate a possible reddening-related systematic in more detail. A detailed comparison with $\alpha_\mathrm{Fe}$ values in \citet[their Table\,9]{CruzReyes2023} was foregone due to the limited metallicity range considered there (MW cluster Cepheids and LMC only), although their observational results generally agree with our predictions.

Figure \ref{fig:logP0_gamma} illustrates the behavior of the wavelength dependence of $\alpha_{\mathrm{M}}$ on pivot period ranging from $0.0 \le \log P_{\rm 0} \le 1.5$.
Low values of $\log P_{\rm 0} = 0.0$ yield a nearly constant $\alpha_{\mathrm{M}}$ with wavelength. However, the wavelength-dependence becomes increasingly stronger at higher $\log P_{\rm 0}$, exhibiting a positive trend as shown in Fig.\,\ref{fig:lambda_gamma}. Last, but not least, we note that the shortest wavelengths yield positive values for $\alpha_\mathrm{M}$ for $\log{P_0} \gtrsim 0.8$. Positive intercept-metallicity effects at $\lambda=1\mu$m occur for $\log{P_0} \gtrsim 1.3$, and $\alpha_\mathrm{M}<0$ for $1/\lambda \lesssim 0.7$, irrespective of $\log{P_0}$.

\begin{figure*}
    \centering
    \includegraphics[width=\hsize]{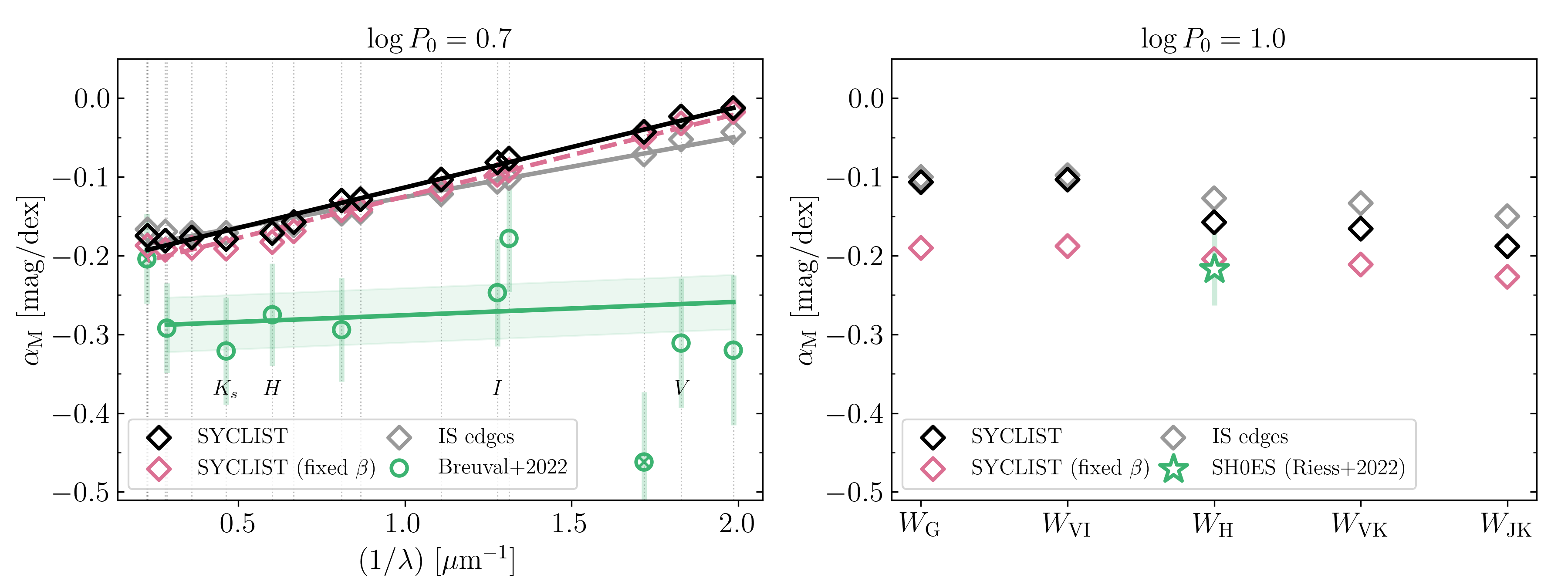}
	\caption{Metallicity effect on the LL intercept, $\alpha_{\mathrm{M}}$, as defined in Eq.\,(\ref{eq:plz_full}). The left panel shows $\alpha_{\rm M}$ as a function of the effective wavelength $1/\lambda$ for a pivot period of $\log P_{\rm 0} = 0.7$, while the right panel shows the comparison for Wesenheit magnitudes and for $\log P_{\rm 0} = 1.0$. Our predictions based on \texttt{SYCLIST}, letting the LL slope vary or be fixed to that of the LMC, and IS edges are shown in black, pink, and grey, respectively. Observational results from \citetalias{Breuval2022} and SH0ES \citepalias{Riess2022} are shown in green in the left and right panels, respectively. In the left panel, the vertical dotted lines show the location of all the photometric passbands considered, and we label some of them to orient the reader.}
	\label{fig:lambda_gamma}
\end{figure*}

\subsubsection{Implications for the Hubble constant\label{sec:H0}}
As mentioned in Sect. \ref{sec:intro}, the impact of Cepheid metallicity on the measurement of \Hcst is largely mitigated by the rather small range of oxygen abundances among Cepheids in the high-mass spiral galaxies hosting type-Ia supernovae that is fully contained within the range of abundances of Cepheids in the host galaxies, notably the MW, the LMC, and NGC\,4258. Correcting differences in Cepheid luminosity due to chemical composition thus mainly improves accuracy, rather than shifting the centre value, cf. Sect.\,6.6 in \citetalias{Riess2022}. However, the intercept-metallicity effect of $\alpha_\mathrm{M} = -0.217 \pm 0.046$ measured as a free parameter of the SH0ES distance ladder \citep[their parameter $\gamma$]{Riess2022} provides the arguably strongest empirical constraint for our theoretical predictions. This measurement is based on thousands of Cepheids across three anchor and 37 SN-host galaxies, relies exclusively on exquisite and homogeneous {\it HST} photometry, which is both insensitive to reddening thanks to infrared photometry and reddening-free by construction thanks to the use of Wesenheit magnitudes. It further assumes a universal LL slope $\beta = -3.299 \pm 0.015$, which is heavily informed by LMC Cepheids, and $\log{P_0}=1.0$.

Results based on \texttt{SYCLIST} populations adopting a fixed LL slope ($\beta$) derived at Z=0.006 and using $\log{P_0}=1.0$ provide the closest correspondence to the observational setup of the SH0ES distance ladder. The right panel of Fig.\,\ref{fig:lambda_gamma} illustrates these results for several Wesenheit magnitudes. Adopting $H-$band Wesenheit magnitudes as used in the SH0ES distance ladder (Eq.\,\ref{eq:SH0ES_mWH}) yields $\alpha_\mathrm{M}=-0.204 \pm 0.030$\,mag\,dex$^{-1}$, in excellent agreement with \citetalias{Riess2022}. Fixing $\beta$ to the Z=0.006 simulations reduces the predicted values for $\alpha_\mathrm{M}$ by $0.05-0.11$\,mag\,dex$^{-1}$, depending on the Wesenheit magnitude considered. Tables \ref{table:plz_syclist} and \ref{table:gamma_IS} report the predicted values for $\alpha_{\mathrm{M}}$. 

\begin{table*}[]
    \caption{Metallicity parameters $\beta_{\mathrm{M}}$ and $\alpha_{\mathrm{M}}$ predicted with \texttt{SYCLIST} populations, either letting the LL slope vary or be fixed to the LMC value (last two columns), for different pivot periods: $\log P_{\rm 0} = 0.0$, 0.7, and 1.0.}
    \label{table:plz_syclist}
    \centering
    \small
    \begin{tabular}{c|cc|cccccc|cc}
\toprule
Band & $\beta_{\mathrm{M}}$ &  $\sigma_{\rm \beta_{\mathrm{M}}}$ & $\alpha_{\mathrm{M}}$ &  $\sigma_{\rm \alpha_{\mathrm{M}}}$ &  $\alpha_{\mathrm{M}}$ &   $\sigma_{\rm \alpha_{\mathrm{M}}}$ &  $\alpha_{\mathrm{M}}$ &   $\sigma_{\rm \alpha_{\mathrm{M}}}$ & $\alpha_{\mathrm{M,fixed\ \beta}}$ &   $\sigma_{\mathrm{\alpha_{\rm M,fixed\ \beta}}}$\\
     &  & & [0.0]  &  [0.0]   &  [0.7]  &  [0.7]   & [1.0]   &    [1.0]  &   & \\
\midrule
 $G_{\rm BP}$ & 0.372 & 0.158 &   $-$0.273 &     0.130 &   $-$0.013 &     0.020 &    0.099 &     0.028 &    $-$0.017 &      0.006 \\
   $V$ & 0.377 & 0.149 &  $-$0.287 &     0.125 &   $-$0.023 &     0.021 &    0.090 &     0.024 &    $-$0.032 &      0.006 \\
   $G$ & 0.320 & 0.128 &  $-$0.267 &     0.113 &   $-$0.043 &     0.023 &    0.053 &     0.015 &    $-$0.049 &      0.001 \\
 $G_{\rm RP}$ & 0.305 & 0.096 &  $-$0.290 &     0.095 &   $-$0.077 &     0.028 &    0.015 &     0.001 &    $-$0.092 &      0.004 \\
   $I$ & 0.302 & 0.092 &  $-$0.293 &     0.093 &   $-$0.082 &     0.028 &    0.009 &     0.001 &    $-$0.097 &      0.004 \\
F090W & 0.266 & 0.086 &  $-$0.289 &     0.091 &   $-$0.103 &     0.030 &   $-$0.023 &     0.005 &    $-$0.116 &      0.010 \\
F115W & 0.226 & 0.065 &  $-$0.287 &     0.079 &   $-$0.128 &     0.033 &   $-$0.060 &     0.014 &    $-$0.141 &      0.015 \\
   $J$ & 0.208 & 0.059 &  $-$0.276 &     0.075 &   $-$0.130 &     0.034 &   $-$0.067 &     0.016 &    $-$0.142 &      0.017 \\
F150W & 0.171 & 0.043 &  $-$0.277 &     0.067 &   $-$0.157 &     0.037 &   $-$0.106 &     0.024 &    $-$0.169 &      0.022 \\
   $H$ & 0.148 & 0.033 & $-$0.275 &     0.062 &   $-$0.171 &     0.038 &   $-$0.127 &     0.028 &    $-$0.183 &      0.026 \\
  $K_s$ & 0.139 & 0.027 & $-$0.276 &     0.058 &   $-$0.179 &     0.039 &   $-$0.137 &     0.031 &    $-$0.191 &      0.027 \\
F277W & 0.148 & 0.028 &  $-$0.280 &     0.058 &   $-$0.177 &     0.039 &   $-$0.133 &     0.031 &    $-$0.190 &      0.026 \\
F356W & 0.127 & 0.024 &  $-$0.270 &     0.056 &   $-$0.181 &     0.039 &   $-$0.143 &     0.032 &    $-$0.192 &      0.028 \\
F444W & 0.141 & 0.028 &  $-$0.274 &     0.058 &   $-$0.175 &     0.039 &   $-$0.132 &     0.030 &    $-$0.187 &      0.026 \\
  \hline
  $W_{\rm G}$ &  0.193 & 0.011 & $-$0.300 &     0.046 &   $-$0.165 &     0.039 &   $-$0.107 &     0.036 &    $-$0.190 &      0.019 \\
 $W_{\rm VI}$ &  0.197 & 0.013 & $-$0.300 &     0.048 &   $-$0.162 &     0.039 &   $-$0.103 &     0.035 &    $-$0.188 &      0.019 \\
  $W_{\rm H}$ &  0.115 & 0.016 & $-$0.273 &     0.052 &   $-$0.192 &     0.041 &   $-$0.158 &     0.036 &    $-$0.204 &      0.030 \\
 $W_{\rm VK}$ &  0.109 & 0.011 & $-$0.275 &     0.049 &   $-$0.198 &     0.041 &   $-$0.166 &     0.038 &    $-$0.211 &      0.031 \\
 $W_{\rm JK}$ &  0.088 & 0.003 & $-$0.276 &     0.045 &   $-$0.215 &     0.043 &   $-$0.188 &     0.042 &    $-$0.227 &      0.034 \\ 
 \bottomrule
\end{tabular}
\tablefoot{NB: Our models do not predict many $1$\,d fundamental-mode Cepheids, cf. Fig.\,\ref{fig:logP_distribution}. Hence, results at $\log P_{\rm 0}=0.0$ are an extrapolation, and other pivot periods should be preferentially considered.}
\end{table*}

\begin{figure}
	\centering
	\includegraphics[width=\hsize]{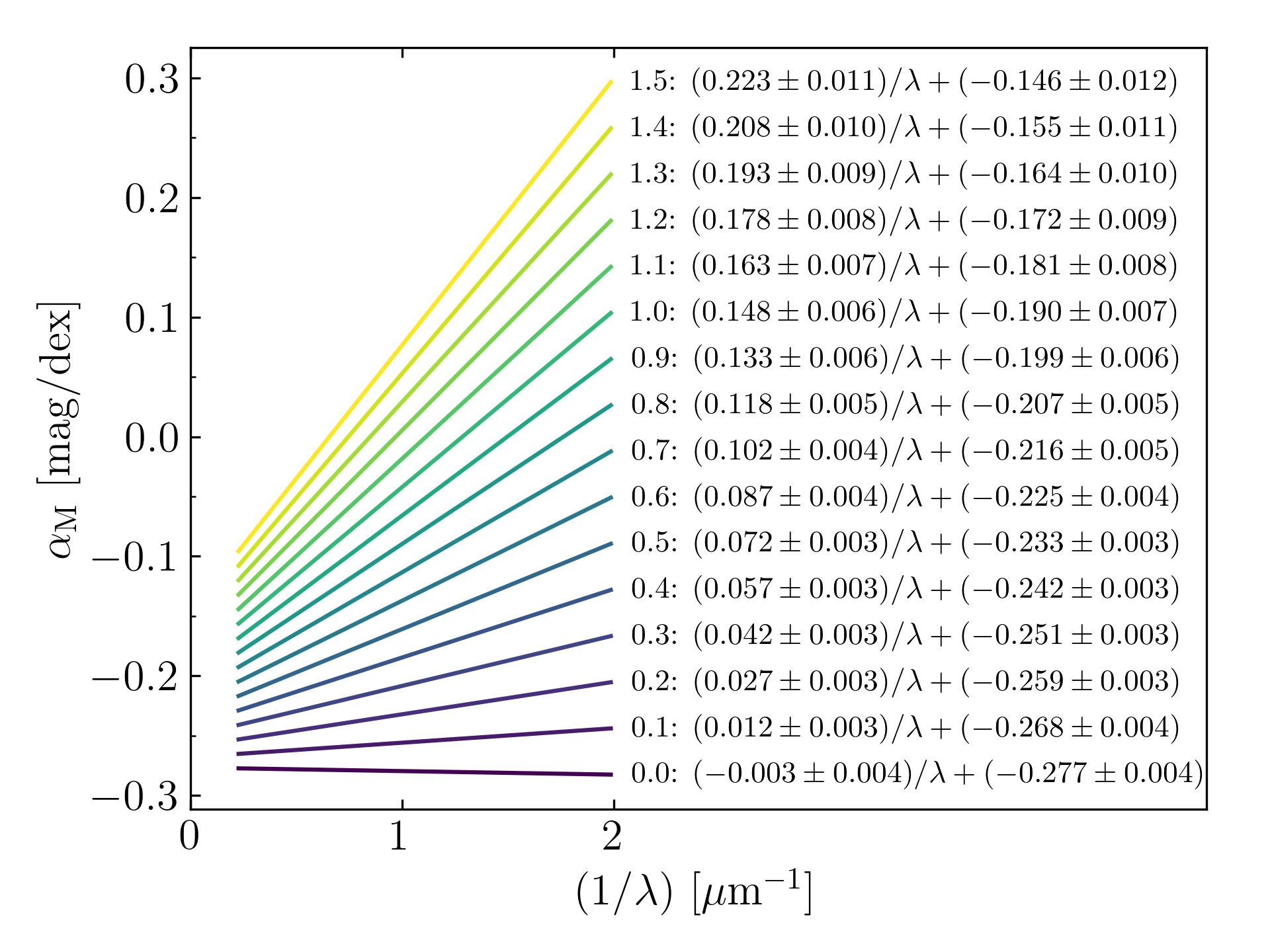}
	\caption{Intercept-metallicity parameter $\alpha_{\mathrm{M}}$ as a function of the effective wavelength $1/\lambda$ for pivot periods ranging between $\log P_{\rm 0}=0.0$ (purple) and 1.5 (yellow), with a step of 0.1. The corresponding fits, based on \texttt{SYCLIST} populations, are annotated next to each line.}
	\label{fig:logP0_gamma}
\end{figure}

\section{Summary and conclusions}
\label{sec:conclusions}
We investigated the effect of chemical composition on the LL using synthetic populations of Cepheids computed using Geneva stellar evolution models, a pulsational instability analysis thereof, and the \texttt{SYCLIST} tool. We note that Geneva models are calibrated to reproduce properties of massive main sequence stars and the Sun, not of evolved stars, such as Cepheids. All models in the initial grids \citep{Ekstrom2012,Georgy2013} have been self-consistently computed as stellar tracks from the zero-age MS to well beyond core He burning.  The synthetic populations computed using \texttt{SYCLIST} interpolate in mass and rotation rate at the original (fixed) metallicity at which the models were computed, with $Z=0.014,\ 0.006,\ \mathrm{ and }\ 0.002$ representing the abundances of the Sun and intermediate-age stars in the LMC and SMC, respectively. Photometric contributions by companion stars were included to maximise realism. 

We tested all predictions using observational constraints from the recent literature. Specifically, we demonstrated that our predictions closely reproduce observed IS boundaries, period distributions, LL slopes, and the intrinsic dispersion of the LL as a function of wavelength and metallicity. Additionally, we note that \citetalias{Anderson2016} previously demonstrated excellent agreement between predictions made by these same models and additional Cepheid properties, including mass-luminosity relations, period-radius relations, rates of period change, and others. This extensive comparison demonstrates the high predictive power of this model set \citep{Ekstrom2012,Georgy2013} for Cepheids.

Our simulations predict LL slopes, $\beta$, to depend on metallicity. This interesting property of the Cepheid LL was previously predicted by \citetalias{Anderson2016} and has been corroborated by another set of models \citep{DeSomma2022}. Section\,\ref{sec:LL_slope_metallicity} shows that observational studies confirm both the direction and magnitude of this effect, despite the measurement being challenging. We note that distance measurements based on Cepheids generally assume universal LL slopes and that further study is needed to test this important prediction. We further caution that the slope-metallicity dependence introduces a dependence of the intercept-metallicity effect on pivot period ($P_0$), which must be explicitly considered in comparisons among studies. 

We found excellent agreement in the intercept-metallicity dependence ($\alpha_\mathrm{M}$) determined in analogy with the SH0ES distance ladder \citepalias{Riess2022}, which provides the most stringent and direct empirical constraint to date, cf. Sect.\,\ref{sec:H0}. We therefore conclude that a) the Cepheid metallicity effect cannot bias the \Hcst measurement by design of the SH0ES distance ladder and that b) our stellar models fully support the approach taken by \citetalias{Riess2022}. Comparisons with other empirical studies from the literature yield the best agreement for studies targeting low-extinction MW Cepheids with accurate parallaxes alongside LMC and SMC Cepheids that assume the well known detached eclipsing binary distances \citep{Pietrzynski2019,Graczyk2020}. However, disagreements at the level of $0.1-0.2$\,mag\,dex$^{-1}$ become more significant at shorter wavelengths (Sect.\,\ref{sec:LL_intercept_metallicity}), suggesting a possible origin in reddening-related systematics. 

Further work is needed to understand the veracity and implications of the LL slope-metallicity dependence as well as a possible non-linearity in the intercept-metallicity dependence. Super-solar metallicity models will be particularly useful to this end, as will improved parallaxes and astrometric systematics in the fourth {\it Gaia} data release scheduled for 2026.

\begin{acknowledgements}
RIA \& SK are funded by the SNSF through a Swiss National Science Foundation Eccellenza Professorial Fellowship (award PCEFP2\_194638). 
\end{acknowledgements}
\bibliographystyle{aa}
\bibliography{literature}

\onecolumn
\begin{appendix}

\section{Terminology of metallicity-dependence}\label{app:PLZ}

Following the terminology used in the current paper, the general form of a LL is $\alpha + \beta \cdot \log{(P/P_0)}$, with pivot period $P_0$. Higher orders in $\log{P}$ as used, e.g., for Miras would then follow with $\gamma \log{P}^2$. In principle, $\alpha$ and $\beta$ can both be metal dependent, and this can be written as $\alpha = \alpha_0 + \alpha_{\mathrm{Fe}}\mathrm{[Fe/H]}$ to indicate an effect related to iron abundance, and analogously for $\beta$. The benefits of using this nomenclature is that a) subscript $_0$ always identifies the fiducial quantity for slope, intercept, and pivot period; b) subscripts can be used to distinguish between different effects/elements considered; c) it is immediately clear which metallicity dependence is discussed, which is more difficult if $\beta = \delta + \gamma\mathrm{[Fe/H]}$, for instance. Table \ref{tab:terminology} provides the equivalence between our terminology in Eq.\,(\ref{eq:plz_full}) and those used in the recent literature. 

\begin{table}[h!]
    \caption{Equivalence for the terminology in the LL parameters between our work and the recent literature.}
    \label{tab:terminology}
    \centering
    \small
    \begin{tabular}{c|c|c|c}
         \toprule
         study & intercept & slope & $\log P_{\rm 0}$\\ 
         \midrule
         this work & $\alpha = \alpha_{\rm 0} + \alpha_{\rm M}\cdot\mh$ & $\beta = \beta_{\rm 0} + \beta_{\rm Z}\cdot\mh$ & varies \\
         \midrule
         S15 & $\beta$ & $\alpha$ & 0.0\\
         \midrule
         B22 & $\beta = \delta + \gamma\cdot\feh$ & $\alpha$ & 0.7 \\
         \midrule
         R22 & $M_{H\rm,1}^W + Z_{W}\cdot\oh$ & $b_{W}$ & 1.0 \\
         \midrule
         T24 & $\alpha + \gamma\cdot\feh$ & $\beta + \delta\cdot\feh$ & 1.0 \\
         \bottomrule
    \end{tabular}
    \tablefoot{S15: \citet{Soszynski2015}, B22: \citet{Breuval2022}, R22: \citet{Riess2022}, T24: \citet{Trentin2024}.}
\end{table}

\clearpage

\section{Leavitt laws parameters for single-wavelength passbands and Wesenheit magnitudes}
\label{app:PLRs}

\begin{table*}[h!]
    \caption{LL slope and intercepts for the \texttt{SYCLIST} populations of IS stars, for different pivot periods: $\log P_{\rm 0} = 0.0$, 0.7, and 1.0.}
    \label{table:PLR_SYCLIST}
    \centering
    \small
    \begin{tabular}{c|c|cc|cccccc}
\toprule
Band &  $Z$  &  $\beta_{\rm SYCLIST}$ & $\sigma_{\beta_{\rm SYCLIST}}$ &  $\alpha_{\rm SYCLIST}$ &  $\sigma_{\alpha_{\rm SYCLIST}}$ &  $\alpha_{\rm SYCLIST}$ &  $\sigma_{\alpha_{\rm SYCLIST}}$ &  $\alpha_{\rm SYCLIST}$ &  $\sigma_{\alpha_{\rm SYCLIST}}$ \\
              &       &        &        &  [0.0] &  [0.0]    &  [0.7] &  [0.7]    & [1.0]  & [1.0]  \\
 \midrule
              & 0.014 & $-$2.278 & 0.008 &  $-$1.442 &    0.007 &  $-$3.037 &    0.003 &  $-$3.720 &    0.003 \\
 $G_{\rm BP}$ & 0.006 & $-$2.535 & 0.007 &  $-$1.243 &    0.005 &  $-$3.017 &    0.002 &  $-$3.778 &    0.003 \\
              & 0.002 & $-$2.601 & 0.006 &  $-$1.204 &    0.003 &  $-$3.025 &    0.002 &  $-$3.805 &    0.004 \\
 \hline
               & 0.014 & $-$2.322 & 0.008 &  $-$1.589 &    0.007 &  $-$3.214 &    0.003 &  $-$3.911 &    0.003 \\
           $V$ & 0.006 & $-$2.574 & 0.006 &  $-$1.387 &    0.005 &  $-$3.189 &    0.002 &  $-$3.962 &    0.003 \\
               & 0.002 & $-$2.650 & 0.006 &  $-$1.338 &    0.003 &  $-$3.193 &    0.002 &  $-$3.988 &    0.004 \\
 \hline
               & 0.014 & $-$2.446 & 0.007 &  $-$1.656 &    0.006 &  $-$3.369 &    0.002 &  $-$4.103 &    0.003 \\
           $G$ & 0.006 & $-$2.662 & 0.006 &  $-$1.472 &    0.004 &  $-$3.335 &    0.002 &  $-$4.134 &    0.002 \\
               & 0.002 & $-$2.725 & 0.005 &  $-$1.424 &    0.002 &  $-$3.331 &    0.002 &  $-$4.149 &    0.003 \\
 \hline
               & 0.014 & $-$2.610 & 0.006 &  $-$2.026 &    0.005 &  $-$3.853 &    0.002 &  $-$4.636 &    0.002 \\
  $G_{\rm RP}$ & 0.006 & $-$2.795 & 0.005 &  $-$1.847 &    0.004 &  $-$3.804 &    0.001 &  $-$4.642 &    0.002 \\
               & 0.002 & $-$2.873 & 0.004 &  $-$1.775 &    0.002 &  $-$3.787 &    0.002 &  $-$4.649 &    0.003 \\
 \hline
               & 0.014 & $-$2.634 & 0.006 &  $-$2.094 &    0.005 &  $-$3.938 &    0.002 &  $-$4.728 &    0.002 \\
           $I$ & 0.006 & $-$2.815 & 0.005 &  $-$1.916 &    0.003 &  $-$3.886 &    0.001 &  $-$4.730 &    0.002 \\
               & 0.002 & $-$2.894 & 0.004 &  $-$1.841 &    0.002 &  $-$3.867 &    0.002 &  $-$4.735 &    0.003 \\
 \hline
               & 0.014 & $-$2.724 & 0.005 &  $-$2.185 &    0.004 &  $-$4.092 &    0.002 &  $-$4.909 &    0.002 \\
         F090W & 0.006 & $-$2.887 & 0.004 &  $-$2.010 &    0.003 &  $-$4.031 &    0.001 &  $-$4.897 &    0.002 \\
               & 0.002 & $-$2.954 & 0.004 &  $-$1.935 &    0.002 &  $-$4.003 &    0.001 &  $-$4.889 &    0.002 \\
 \hline
               & 0.014 & $-$2.854 & 0.004 &  $-$2.345 &    0.003 &  $-$4.342 &    0.001 &  $-$5.198 &    0.002 \\
         F115W & 0.006 & $-$2.987 & 0.003 &  $-$2.179 &    0.003 &  $-$4.269 &    0.001 &  $-$5.165 &    0.001 \\
               & 0.002 & $-$3.049 & 0.003 &  $-$2.097 &    0.001 &  $-$4.232 &    0.001 &  $-$5.146 &    0.002 \\
 \hline
               & 0.014 & $-$2.875 & 0.004 &  $-$2.369 &    0.003 &  $-$4.381 &    0.001 &  $-$5.243 &    0.002 \\
           $J$ & 0.006 & $-$2.996 & 0.003 &  $-$2.210 &    0.003 &  $-$4.308 &    0.001 &  $-$5.206 &    0.001 \\
               & 0.002 & $-$3.054 & 0.003 &  $-$2.131 &    0.001 &  $-$4.269 &    0.001 &  $-$5.185 &    0.002 \\
 \hline
               & 0.014 & $-$3.008 & 0.003 &  $-$2.489 &    0.002 &  $-$4.595 &    0.001 &  $-$5.497 &    0.001 \\
         F150W & 0.006 & $-$3.104 & 0.003 &  $-$2.336 &    0.002 &  $-$4.509 &    0.001 &  $-$5.440 &    0.001 \\
               & 0.002 & $-$3.156 & 0.003 &  $-$2.251 &    0.001 &  $-$4.460 &    0.001 &  $-$5.406 &    0.002 \\
 \hline
               & 0.014 & $-$3.080 & 0.003 &  $-$2.592 &    0.002 &  $-$4.748 &    0.001 &  $-$5.671 &    0.001 \\
           $H$ & 0.006 & $-$3.160 & 0.003 &  $-$2.444 &    0.002 &  $-$4.655 &    0.001 &  $-$5.603 &    0.001 \\
               & 0.002 & $-$3.207 & 0.003 &  $-$2.355 &    0.001 &  $-$4.600 &    0.001 &  $-$5.563 &    0.002 \\
 \hline
               & 0.014 & $-$3.117 & 0.002 &  $-$2.633 &    0.002 &  $-$4.815 &    0.001 &  $-$5.751 &    0.001 \\
         $K_s$ & 0.006 & $-$3.189 & 0.002 &  $-$2.488 &    0.002 &  $-$4.720 &    0.001 &  $-$5.676 &    0.001 \\
               & 0.002 & $-$3.237 & 0.003 &  $-$2.396 &    0.001 &  $-$4.662 &    0.001 &  $-$5.633 &    0.002 \\
 \hline
               & 0.014 & $-$3.104 & 0.002 &  $-$2.652 &    0.002 &  $-$4.825 &    0.001 &  $-$5.756 &    0.001 \\
         F277W & 0.006 & $-$3.180 & 0.003 &  $-$2.504 &    0.002 &  $-$4.730 &    0.001 &  $-$5.684 &    0.001 \\
               & 0.002 & $-$3.231 & 0.003 &  $-$2.411 &    0.001 &  $-$4.673 &    0.001 &  $-$5.642 &    0.002 \\
 \hline
               & 0.014 & $-$3.135 & 0.002 &  $-$2.653 &    0.002 &  $-$4.848 &    0.001 &  $-$5.789 &    0.001 \\
         F356W & 0.006 & $-$3.200 & 0.002 &  $-$2.511 &    0.002 &  $-$4.751 &    0.001 &  $-$5.711 &    0.001 \\
               & 0.002 & $-$3.244 & 0.002 &  $-$2.422 &    0.001 &  $-$4.693 &    0.001 &  $-$5.666 &    0.002 \\
 \hline
               & 0.014 & $-$3.101 & 0.002 &  $-$2.652 &    0.002 &  $-$4.823 &    0.001 &  $-$5.753 &    0.001 \\
         F444W & 0.006 & $-$3.174 & 0.003 &  $-$2.507 &    0.002 &  $-$4.729 &    0.001 &  $-$5.681 &    0.001 \\
               & 0.002 & $-$3.222 & 0.003 &  $-$2.417 &    0.001 &  $-$4.672 &    0.001 &  $-$5.639 &    0.002 \\
 \hline\hline
               & 0.014 & $-$3.078 & 0.003 &  $-$2.766 &    0.002 &  $-$4.920 &    0.001 &  $-$5.844 &    0.001 \\
   $W_{\rm G}$ & 0.006 & $-$3.157 & 0.003 &  $-$2.620 &    0.002 &  $-$4.830 &    0.001 &  $-$5.777 &    0.001 \\
               & 0.002 & $-$3.241 & 0.003 &  $-$2.510 &    0.001 &  $-$4.779 &    0.001 &  $-$5.751 &    0.002 \\
 \hline
               & 0.014 & $-$3.065 & 0.003 &  $-$2.795 &    0.002 &  $-$4.941 &    0.001 &  $-$5.861 &    0.001 \\
  $W_{\rm VI}$ & 0.006 & $-$3.148 & 0.003 &  $-$2.649 &    0.002 &  $-$4.852 &    0.001 &  $-$5.796 &    0.001 \\
               & 0.002 & $-$3.233 & 0.003 &  $-$2.539 &    0.001 &  $-$4.802 &    0.001 &  $-$5.771 &    0.002 \\
 \hline
               & 0.014 & $-$3.188 & 0.002 &  $-$2.793 &    0.002 &  $-$5.025 &    0.001 &  $-$5.981 &    0.001 \\
   $W_{\rm H}$ & 0.006 & $-$3.243 & 0.002 &  $-$2.653 &    0.002 &  $-$4.923 &    0.001 &  $-$5.896 &    0.001 \\
               & 0.002 & $-$3.286 & 0.002 &  $-$2.559 &    0.001 &  $-$4.860 &    0.001 &  $-$5.846 &    0.002 \\
 \hline
               & 0.014 & $-$3.218 & 0.002 &  $-$2.766 &    0.002 &  $-$5.019 &    0.001 &  $-$5.984 &    0.001 \\
  $W_{\rm VK}$ & 0.006 & $-$3.267 & 0.002 &  $-$2.627 &    0.002 &  $-$4.914 &    0.001 &  $-$5.894 &    0.001 \\
               & 0.002 & $-$3.311 & 0.002 &  $-$2.531 &    0.001 &  $-$4.849 &    0.001 &  $-$5.842 &    0.001 \\
 \hline
               & 0.014 & $-$3.295 & 0.002 &  $-$2.828 &    0.001 &  $-$5.135 &    0.001 &  $-$6.123 &    0.001 \\
  $W_{\rm JK}$ & 0.006 & $-$3.330 & 0.002 &  $-$2.692 &    0.002 &  $-$5.023 &    0.001 &  $-$6.022 &    0.001 \\
               & 0.002 & $-$3.370 & 0.002 &  $-$2.592 &    0.001 &  $-$4.951 &    0.001 &  $-$5.962 &    0.001 \\
\bottomrule
    \end{tabular}
\end{table*}

\begin{table*}[h!]
    \caption{LL intercepts for the \texttt{SYCLIST} populations of IS stars, using a slope ($\beta$) fixed to the value for the LMC, for different pivot periods: $\log P_{\rm 0} = 0.0$, 0.7, and 1.0.}
    \label{table:PLR_SYCLIST_fixed}
    \centering
    \small
    \begin{tabular}{c|c|cc|cccccc}
\toprule
Band &  $Z$  &  $\beta_{\rm SYCLIST}$ & $\sigma_{\beta_{\rm SYCLIST}}$ &  $\alpha_{\rm SYCLIST, fixed\ \beta}$ &  $\sigma_{\alpha_{\rm SYCLIST, fixed\ \beta}}$ &  $\alpha_{\rm SYCLIST, fixed\ \beta}$ &  $\sigma_{\alpha_{\rm SYCLIST, fixed\ \beta}}$ &  $\alpha_{\rm SYCLIST, fixed\ \beta}$ &  $\sigma_{\alpha_{\rm SYCLIST, fixed\ \beta}}$ \\
              &       &        &        &  [0.0] &  [0.0]    &  [0.7] &  [0.7]    & [1.0]  & [1.0]  \\
\midrule
              & 0.014 & - & -          & $-$1.244 &    0.001 &  $-$3.019 &    0.001 &  $-$3.779 &    0.001 \\
 $G_{\rm BP}$ & 0.006 & $-$2.535 & 0.007 & $-$1.243 &    0.005 &  $-$3.017 &    0.002 &  $-$3.778 &    0.003 \\
              & 0.002 & - & -          & $-$1.230 &    0.001 &  $-$3.004 &    0.001 &  $-$3.765 &    0.001 \\
 \hline
              & 0.014 & - & -          & $-$1.395 &    0.001 &  $-$3.197 &    0.001 &  $-$3.969 &    0.001 \\
          $V$ & 0.006 & $-$2.574 & 0.006 & $-$1.387 &    0.005 &  $-$3.189 &    0.002 &  $-$3.962 &    0.003 \\
              & 0.002 & - & -          & $-$1.368 &    0.001 &  $-$3.170 &    0.001 &  $-$3.942 &    0.001 \\
   \hline
              & 0.014 & - & -          & $-$1.491 &    0.001 &  $-$3.354 &    0.001 &  $-$4.153 &    0.001 \\
          $G$ & 0.006 & $-$2.662 & 0.006 & $-$1.472 &    0.004 &  $-$3.335 &    0.002 &  $-$4.134 &    0.002 \\
              & 0.002 & - & -          & $-$1.449 &    0.001 &  $-$3.312 &    0.001 &  $-$4.111 &    0.001 \\
   \hline
              & 0.014 & - & -          & $-$1.884 &    0.001 &  $-$3.840 &    0.001 &  $-$4.679 &    0.001 \\
 $G_{\rm RP}$ & 0.006 & $-$2.795 & 0.005 & $-$1.847 &    0.004 &  $-$3.804 &    0.001 &  $-$4.642 &    0.002 \\
              & 0.002 & - & -          & $-$1.806 &    0.001 &  $-$3.763 &    0.001 &  $-$4.601 &    0.001 \\
 \hline
              & 0.014 & - & -          & $-$1.955 &    0.001 &  $-$3.925 &    0.001 &  $-$4.770 &    0.001 \\
          $I$ & 0.006 & $-$2.815 & 0.005 & $-$1.916 &    0.003 &  $-$3.886 &    0.001 &  $-$4.730 &    0.002 \\
              & 0.002 & - & -          & $-$1.872 &    0.001 &  $-$3.843 &    0.001 &  $-$4.687 &    0.001 \\
 \hline
              & 0.014 & - & -          & $-$2.060 &    0.001 &  $-$4.081 &    0.001 &  $-$4.947 &    0.001 \\
        F090W & 0.006 & $-$2.887 & 0.004 & $-$2.010 &    0.003 &  $-$4.031 &    0.001 &  $-$4.897 &    0.002 \\
              & 0.002 & - & -          & $-$1.962 &    0.001 &  $-$3.983 &    0.001 &  $-$4.849 &    0.001 \\
 \hline
              & 0.014 & - & -          & $-$2.242 &    0.001 &  $-$4.333 &    0.001 &  $-$5.229 &    0.001 \\
        F115W & 0.006 & $-$2.987 & 0.003 & $-$2.179 &    0.003 &  $-$4.269 &    0.001 &  $-$5.165 &    0.001 \\
              & 0.002 & - & -          & $-$2.122 &    0.001 &  $-$4.213 &    0.001 &  $-$5.109 &    0.001 \\
 \hline
              & 0.014 & - & -          & $-$2.275 &    0.001 &  $-$4.373 &    0.001 &  $-$5.272 &    0.001 \\
          $J$ & 0.006 & $-$2.996 & 0.003 & $-$2.210 &    0.003 &  $-$4.308 &    0.001 &  $-$5.206 &    0.001 \\
              & 0.002 & - & -          & $-$2.154 &    0.001 &  $-$4.251 &    0.001 &  $-$5.150 &    0.001 \\
 \hline
              & 0.014 & - & -          & $-$2.415 &    0.001 &  $-$4.588 &    0.001 &  $-$5.519 &    0.001 \\
        F150W & 0.006 & $-$3.104 & 0.003 & $-$2.336 &    0.002 &  $-$4.509 &    0.001 &  $-$5.440 &    0.001 \\
              & 0.002 & - & -          & $-$2.271 &    0.001 &  $-$4.444 &    0.001 &  $-$5.375 &    0.001 \\
 \hline
              & 0.014 & - & -          & $-$2.530 &    0.001 &  $-$4.742 &    0.001 &  $-$5.690 &    0.001 \\
          $H$ & 0.006 & $-$3.160 & 0.003 & $-$2.444 &    0.002 &  $-$4.655 &    0.001 &  $-$5.603 &    0.001 \\
              & 0.002 & - & -          & $-$2.374 &    0.001 &  $-$4.586 &    0.001 &  $-$5.534 &    0.001 \\
 \hline
              & 0.014 & - & -          & $-$2.578 &    0.001 &  $-$4.811 &    0.001 &  $-$5.767 &    0.001 \\
        $K_s$ & 0.006 & $-$3.189 & 0.002 & $-$2.488 &    0.002 &  $-$4.720 &    0.001 &  $-$5.676 &    0.001 \\
              & 0.002 & - & -          & $-$2.415 &    0.001 &  $-$4.647 &    0.001 &  $-$5.604 &    0.001 \\
 \hline
              & 0.014 & - & -          & $-$2.594 &    0.001 &  $-$4.820 &    0.001 &  $-$5.774 &    0.001 \\
        F277W & 0.006 & $-$3.180 & 0.003 & $-$2.504 &    0.002 &  $-$4.730 &    0.001 &  $-$5.684 &    0.001 \\
              & 0.002 & - & -          & $-$2.432 &    0.001 &  $-$4.658 &    0.001 &  $-$5.611 &    0.001 \\
 \hline
              & 0.014 & - & -          & $-$2.603 &    0.001 &  $-$4.844 &    0.001 &  $-$5.804 &    0.001 \\
        F356W & 0.006 & $-$3.200 & 0.002 & $-$2.511 &    0.002 &  $-$4.751 &    0.001 &  $-$5.711 &    0.001 \\
              & 0.002 & - & -          & $-$2.439 &    0.001 &  $-$4.679 &    0.001 &  $-$5.639 &    0.001 \\
 \hline
              & 0.014 & - & -          & $-$2.596 &    0.001 &  $-$4.817 &    0.001 &  $-$5.770 &    0.001 \\
        F444W & 0.006 & $-$3.174 & 0.003 & $-$2.507 &    0.002 &  $-$4.729 &    0.001 &  $-$5.681 &    0.001 \\
              & 0.002 & - & -          & $-$2.436 &    0.001 &  $-$4.658 &    0.001 &  $-$5.610 &    0.001 \\
 \hline\hline
              & 0.014 & - & -          & $-$2.705 &    0.001 &  $-$4.915 &    0.001 &  $-$5.862 &    0.001 \\
  $W_{\rm G}$ & 0.006 & $-$3.157 & 0.003 & $-$2.620 &    0.002 &  $-$4.830 &    0.001 &  $-$5.777 &    0.001 \\
              & 0.002 & - & -          & $-$2.543 &    0.001 &  $-$4.753 &    0.001 &  $-$5.700 &    0.001 \\
  \hline
              & 0.014 & - & -          & $-$2.732 &    0.001 &  $-$4.936 &    0.001 &  $-$5.880 &    0.001 \\
 $W_{\rm VI}$ & 0.006 & $-$3.148 & 0.003 & $-$2.649 &    0.002 &  $-$4.852 &    0.001 &  $-$5.796 &    0.001 \\
              & 0.002 & - & -          & $-$2.572 &    0.001 &  $-$4.776 &    0.001 &  $-$5.720 &    0.001 \\
 \hline
              & 0.014 & - & -          & $-$2.751 &    0.001 &  $-$5.021 &    0.001 &  $-$5.994 &    0.001 \\
  $W_{\rm H}$ & 0.006 & $-$3.243 & 0.002 & $-$2.653 &    0.002 &  $-$4.923 &    0.001 &  $-$5.896 &    0.001 \\
              & 0.002 & - & -          & $-$2.577 &    0.001 &  $-$4.847 &    0.001 &  $-$5.819 &    0.001 \\
  \hline
              & 0.014 & - & -          & $-$2.729 &    0.001 &  $-$5.016 &    0.001 &  $-$5.996 &    0.001 \\
 $W_{\rm VK}$ & 0.006 & $-$3.267 & 0.002 & $-$2.627 &    0.002 &  $-$4.914 &    0.001 &  $-$5.894 &    0.001 \\
              & 0.002 & - & -          & $-$2.548 &    0.001 &  $-$4.835 &    0.001 &  $-$5.815 &    0.001 \\
 \hline
              & 0.014 & - & -          & $-$2.801 &    0.001 &  $-$5.132 &    0.001 &  $-$6.131 &    0.001 \\
 $W_{\rm JK}$ & 0.006 & $-$3.330 & 0.002 & $-$2.692 &    0.002 &  $-$5.023 &    0.001 &  $-$6.022 &    0.001 \\
              & 0.002 & - & -          & $-$2.607 &    0.001 &  $-$4.939 &    0.001 &  $-$5.938 &    0.001 \\
\bottomrule
\end{tabular}
\end{table*}

\begin{table*}[h!]
    \caption{LL slope and intercepts for the blue edge of the IS, for different pivot periods: $\log P_{\rm 0} = 0.0$, 0.7, and 1.0.}
    \label{table:PLR_ISedges_blue}
    \centering
    \small
    \begin{tabular}{c|c|cc|cccccc}
\toprule
Band &  $Z$ & $\beta_{\rm blue}$ &  $\sigma_{\beta_{\rm blue}}$ &  $\alpha_{\rm blue}$ &  $\sigma_{\alpha_{\rm blue}}$ &  $\alpha_{\rm blue}$ &  $\sigma_{\alpha_{\rm blue}}$ &  $\alpha_{\rm blue}$ &  $\sigma_{\alpha_{\rm blue}}$ \\
              &       &        &       &  [0.0]  &  [0.0]   &  [0.7]  &  [0.7]   & [1.0]   &    [1.0] \\
\midrule
              & 0.014 & $-$2.785 & 0.046 &  $-$1.520 &    0.056 &  $-$3.470 &    0.029 &  $-$4.305 &    0.023 \\
 $G_{\rm BP}$ & 0.006 & $-$2.757 & 0.045 &  $-$1.646 &    0.064 &  $-$3.576 &    0.036 &  $-$4.403 &    0.026 \\
              & 0.002 & $-$2.951 & 0.031 &  $-$1.408 &    0.038 &  $-$3.474 &    0.021 &  $-$4.359 &    0.018 \\
 \hline
              & 0.014 & $-$2.822 & 0.045 &  $-$1.651 &    0.055 &  $-$3.627 &    0.028 &  $-$4.473 &    0.022 \\
          $V$ & 0.006 & $-$2.791 & 0.044 &  $-$1.774 &    0.063 &  $-$3.727 &    0.035 &  $-$4.565 &    0.026 \\
              & 0.002 & $-$2.986 & 0.031 &  $-$1.532 &    0.037 &  $-$3.622 &    0.021 &  $-$4.518 &    0.018 \\
 \hline
              & 0.014 & $-$2.888 & 0.043 &  $-$1.723 &    0.052 &  $-$3.745 &    0.027 &  $-$4.611 &    0.021 \\
          $G$ & 0.006 & $-$2.850 & 0.043 &  $-$1.842 &    0.062 &  $-$3.837 &    0.035 &  $-$4.692 &    0.025 \\
              & 0.002 & $-$3.039 & 0.031 &  $-$1.603 &    0.037 &  $-$3.731 &    0.021 &  $-$4.643 &    0.018 \\
 \hline
              & 0.014 & $-$3.004 & 0.041 &  $-$2.061 &    0.049 &  $-$4.164 &    0.025 &  $-$5.065 &    0.020 \\
 $G_{\rm RP}$ & 0.006 & $-$2.958 & 0.042 &  $-$2.169 &    0.060 &  $-$4.240 &    0.034 &  $-$5.127 &    0.025 \\
              & 0.002 & $-$3.147 & 0.030 &  $-$1.922 &    0.037 &  $-$4.125 &    0.021 &  $-$5.069 &    0.017 \\
 \hline
              & 0.014 & $-$3.021 & 0.040 &  $-$2.125 &    0.048 &  $-$4.239 &    0.025 &  $-$5.146 &    0.020 \\
          $I$ & 0.006 & $-$2.973 & 0.042 &  $-$2.232 &    0.060 &  $-$4.313 &    0.033 &  $-$5.205 &    0.024 \\
              & 0.002 & $-$3.162 & 0.030 &  $-$1.983 &    0.037 &  $-$4.197 &    0.021 &  $-$5.145 &    0.017 \\
 \hline
              & 0.014 & $-$3.075 & 0.038 &  $-$2.201 &    0.046 &  $-$4.353 &    0.024 &  $-$5.276 &    0.019 \\
        F090W & 0.006 & $-$3.019 & 0.041 &  $-$2.305 &    0.059 &  $-$4.419 &    0.033 &  $-$5.324 &    0.024 \\
              & 0.002 & $-$3.198 & 0.031 &  $-$2.065 &    0.037 &  $-$4.303 &    0.021 &  $-$5.263 &    0.017 \\
 \hline
              & 0.014 & $-$3.154 & 0.037 &  $-$2.348 &    0.044 &  $-$4.556 &    0.023 &  $-$5.502 &    0.018 \\
        F115W & 0.006 & $-$3.090 & 0.041 &  $-$2.447 &    0.059 &  $-$4.610 &    0.033 &  $-$5.537 &    0.024 \\
              & 0.002 & $-$3.264 & 0.031 &  $-$2.207 &    0.037 &  $-$4.492 &    0.021 &  $-$5.471 &    0.018 \\
 \hline
              & 0.014 & $-$3.156 & 0.036 &  $-$2.383 &    0.044 &  $-$4.593 &    0.023 &  $-$5.539 &    0.018 \\
          $J$ & 0.006 & $-$3.091 & 0.041 &  $-$2.482 &    0.059 &  $-$4.646 &    0.033 &  $-$5.573 &    0.024 \\
              & 0.002 & $-$3.264 & 0.031 &  $-$2.242 &    0.037 &  $-$4.527 &    0.021 &  $-$5.506 &    0.018 \\
 \hline
              & 0.014 & $-$3.246 & 0.035 &  $-$2.481 &    0.042 &  $-$4.754 &    0.022 &  $-$5.728 &    0.017 \\
        F150W & 0.006 & $-$3.172 & 0.041 &  $-$2.575 &    0.058 &  $-$4.796 &    0.033 &  $-$5.747 &    0.024 \\
              & 0.002 & $-$3.337 & 0.031 &  $-$2.339 &    0.038 &  $-$4.675 &    0.021 &  $-$5.676 &    0.018 \\
 \hline
              & 0.014 & $-$3.290 & 0.034 &  $-$2.578 &    0.041 &  $-$4.881 &    0.021 &  $-$5.867 &    0.017 \\
          $H$ & 0.006 & $-$3.211 & 0.041 &  $-$2.669 &    0.059 &  $-$4.916 &    0.033 &  $-$5.880 &    0.024 \\
              & 0.002 & $-$3.372 & 0.031 &  $-$2.434 &    0.038 &  $-$4.795 &    0.021 &  $-$5.806 &    0.018 \\
 \hline
              & 0.014 & $-$3.314 & 0.034 &  $-$2.614 &    0.040 &  $-$4.934 &    0.021 &  $-$5.928 &    0.016 \\
        $K_s$ & 0.006 & $-$3.233 & 0.041 &  $-$2.703 &    0.059 &  $-$4.966 &    0.033 &  $-$5.936 &    0.024 \\
              & 0.002 & $-$3.393 & 0.031 &  $-$2.468 &    0.038 &  $-$4.844 &    0.021 &  $-$5.861 &    0.018 \\
 \hline
              & 0.014 & $-$3.308 & 0.034 &  $-$2.631 &    0.041 &  $-$4.947 &    0.021 &  $-$5.940 &    0.016 \\
        F277W & 0.006 & $-$3.228 & 0.041 &  $-$2.720 &    0.058 &  $-$4.980 &    0.033 &  $-$5.948 &    0.024 \\
              & 0.002 & $-$3.390 & 0.031 &  $-$2.484 &    0.038 &  $-$4.857 &    0.021 &  $-$5.874 &    0.018 \\
 \hline
              & 0.014 & $-$3.321 & 0.033 &  $-$2.638 &    0.040 &  $-$4.962 &    0.021 &  $-$5.959 &    0.016 \\
        F356W & 0.006 & $-$3.239 & 0.041 &  $-$2.726 &    0.059 &  $-$4.994 &    0.033 &  $-$5.966 &    0.024 \\
              & 0.002 & $-$3.398 & 0.031 &  $-$2.492 &    0.038 &  $-$4.871 &    0.021 &  $-$5.890 &    0.018 \\
 \hline
              & 0.014 & $-$3.301 & 0.034 &  $-$2.638 &    0.041 &  $-$4.949 &    0.021 &  $-$5.939 &    0.017 \\
        F444W & 0.006 & $-$3.222 & 0.041 &  $-$2.728 &    0.058 &  $-$4.983 &    0.033 &  $-$5.949 &    0.024 \\
              & 0.002 & $-$3.383 & 0.031 &  $-$2.491 &    0.038 &  $-$4.860 &    0.021 &  $-$5.875 &    0.018 \\
 \hline\hline
              & 0.014 & $-$3.304 & 0.035 &  $-$2.751 &    0.042 &  $-$5.064 &    0.022 &  $-$6.055 &    0.017 \\
  $W_{\rm G}$ & 0.006 & $-$3.233 & 0.041 &  $-$2.836 &    0.059 &  $-$5.099 &    0.033 &  $-$6.069 &    0.024 \\
              & 0.002 & $-$3.412 & 0.031 &  $-$2.579 &    0.038 &  $-$4.967 &    0.021 &  $-$5.991 &    0.018 \\
 \hline
              & 0.014 & $-$3.297 & 0.035 &  $-$2.781 &    0.042 &  $-$5.089 &    0.022 &  $-$6.078 &    0.017 \\
 $W_{\rm VI}$ & 0.006 & $-$3.226 & 0.041 &  $-$2.867 &    0.059 &  $-$5.125 &    0.033 &  $-$6.093 &    0.024 \\
              & 0.002 & $-$3.405 & 0.031 &  $-$2.610 &    0.038 &  $-$4.994 &    0.021 &  $-$6.015 &    0.018 \\
 \hline
              & 0.014 & $-$3.356 & 0.033 &  $-$2.769 &    0.040 &  $-$5.118 &    0.021 &  $-$6.125 &    0.016 \\
  $W_{\rm H}$ & 0.006 & $-$3.270 & 0.041 &  $-$2.856 &    0.059 &  $-$5.145 &    0.033 &  $-$6.126 &    0.024 \\
              & 0.002 & $-$3.427 & 0.032 &  $-$2.622 &    0.038 &  $-$5.021 &    0.021 &  $-$6.048 &    0.018 \\
 \hline
              & 0.014 & $-$3.377 & 0.033 &  $-$2.736 &    0.040 &  $-$5.100 &    0.021 &  $-$6.113 &    0.016 \\
 $W_{\rm VK}$ & 0.006 & $-$3.289 & 0.041 &  $-$2.821 &    0.059 &  $-$5.124 &    0.033 &  $-$6.110 &    0.024 \\
              & 0.002 & $-$3.445 & 0.032 &  $-$2.587 &    0.038 &  $-$4.999 &    0.022 &  $-$6.032 &    0.018 \\
 \hline
              & 0.014 & $-$3.431 & 0.032 &  $-$2.784 &    0.039 &  $-$5.185 &    0.020 &  $-$6.214 &    0.016 \\
 $W_{\rm JK}$ & 0.006 & $-$3.338 & 0.041 &  $-$2.865 &    0.059 &  $-$5.202 &    0.033 &  $-$6.203 &    0.024 \\
              & 0.002 & $-$3.488 & 0.032 &  $-$2.635 &    0.039 &  $-$5.076 &    0.022 &  $-$6.122 &    0.018 \\
 \bottomrule
    \end{tabular}
\end{table*}

\begin{sidewaystable*}[h!]
    \caption{LL slope and intercepts for the red edge of the IS, for different pivot periods: $\log P_{\rm 0} = 0.0$, 0.7, and 1.0. Parameters for the edges corresponding to the first crossing, and the second and third crossing averaged together are given separately.}
    \label{table:PLR_ISedges_red}
    \centering
    \tiny
    \begin{tabular}{c|c|cc|cccccc|cc|cccccc}
\toprule
Band &  $Z$ & $\beta_{\rm red,1}$ &  $\sigma_{\beta_{\rm red,1}}$ &  $\alpha_{\rm red,1}$ &  $\sigma_{\alpha_{\rm red,1}}$ &  $\alpha_{\rm red,1}$ &  $\sigma_{\alpha_{\rm red,1}}$ &  $\alpha_{\rm red,1}$ &  $\sigma_{\alpha_{\rm red,1}}$ &  $\beta_{\rm red,2+3}$ &  $\sigma_{\beta_{\rm red,2+3}}$ &  $\alpha_{\rm red,2+3}$ &  $\sigma_{\alpha_{\rm red,2+3}}$ &  $\alpha_{\rm red,2+3}$ &  $\sigma_{\alpha_{\rm red,2+3}}$ &  $\alpha_{\rm red,2+3}$ &  $\sigma_{\alpha_{\rm red,2+3}}$ \\
              &       &        &       &  [0.0]   &  [0.0]    & [0.7]    & [0.7]     & [1.0]    &  [1.0]    &        &        &  [0.0]    &  [0.0]     & [0.7]     & [0.7]      & [1.0]     &  [1.0]    \\
\midrule
              & 0.014 & $-$2.551 & 0.067 &   $-$0.872 &     0.056 &   $-$2.657 &     0.058 &   $-$3.422 &     0.070 & $-$1.778 &  0.100 &    $-$1.514 &      0.137 &    $-$2.759 &      0.076 &    $-$3.292 &      0.056 \\
 $G_{\rm BP}$ & 0.006 & $-$2.745 & 0.045 &   $-$0.838 &     0.032 &   $-$2.759 &     0.039 &   $-$3.582 &     0.049 & $-$2.076 &  0.054 &    $-$1.247 &      0.071 &    $-$2.700 &      0.040 &    $-$3.323 &      0.031 \\
              & 0.002 & $-$2.901 & 0.035 &   $-$0.820 &     0.021 &   $-$2.850 &     0.029 &   $-$3.721 &     0.037 & $-$2.415 &  0.046 &    $-$0.948 &      0.058 &    $-$2.638 &      0.033 &    $-$3.363 &      0.028 \\
 \hline
              & 0.014 & $-$2.602 & 0.067 &   $-$1.027 &     0.056 &   $-$2.849 &     0.058 &   $-$3.629 &     0.069 & $-$1.827 &  0.098 &    $-$1.673 &      0.135 &    $-$2.952 &      0.074 &    $-$3.500 &      0.055 \\
          $V$ & 0.006 & $-$2.790 & 0.044 &   $-$0.993 &     0.032 &   $-$2.946 &     0.039 &   $-$3.784 &     0.048 & $-$2.122 &  0.054 &    $-$1.402 &      0.071 &    $-$2.887 &      0.040 &    $-$3.524 &      0.031 \\
              & 0.002 & $-$2.943 & 0.035 &   $-$0.974 &     0.021 &   $-$3.034 &     0.029 &   $-$3.917 &     0.037 & $-$2.456 &  0.046 &    $-$1.103 &      0.058 &    $-$2.822 &      0.033 &    $-$3.559 &      0.028 \\
 \hline
              & 0.014 & $-$2.709 & 0.060 &   $-$1.155 &     0.050 &   $-$3.052 &     0.052 &   $-$3.865 &     0.062 & $-$2.057 &  0.087 &    $-$1.667 &      0.119 &    $-$3.107 &      0.066 &    $-$3.724 &      0.048 \\
          $G$ & 0.006 & $-$2.875 & 0.041 &   $-$1.118 &     0.030 &   $-$3.130 &     0.036 &   $-$3.992 &     0.044 & $-$2.291 &  0.046 &    $-$1.444 &      0.060 &    $-$3.047 &      0.034 &    $-$3.734 &      0.027 \\
              & 0.002 & $-$3.012 & 0.033 &   $-$1.094 &     0.020 &   $-$3.202 &     0.027 &   $-$4.105 &     0.035 & $-$2.573 &  0.043 &    $-$1.187 &      0.053 &    $-$2.988 &      0.031 &    $-$3.760 &      0.026 \\
 \hline
              & 0.014 & $-$2.879 & 0.056 &   $-$1.577 &     0.046 &   $-$3.592 &     0.048 &   $-$4.456 &     0.057 & $-$2.285 &  0.078 &    $-$2.027 &      0.106 &    $-$3.627 &      0.059 &    $-$4.312 &      0.043 \\
 $G_{\rm RP}$ & 0.006 & $-$3.021 & 0.039 &   $-$1.537 &     0.028 &   $-$3.651 &     0.034 &   $-$4.558 &     0.042 & $-$2.478 &  0.042 &    $-$1.821 &      0.055 &    $-$3.556 &      0.031 &    $-$4.299 &      0.024 \\
              & 0.002 & $-$3.142 & 0.032 &   $-$1.507 &     0.019 &   $-$3.706 &     0.026 &   $-$4.649 &     0.034 & $-$2.725 &  0.041 &    $-$1.585 &      0.051 &    $-$3.493 &      0.029 &    $-$4.310 &      0.025 \\
 \hline
              & 0.014 & $-$2.903 & 0.055 &   $-$1.653 &     0.046 &   $-$3.685 &     0.047 &   $-$4.556 &     0.057 & $-$2.318 &  0.076 &    $-$2.093 &      0.105 &    $-$3.716 &      0.058 &    $-$4.412 &      0.042 \\
          $I$ & 0.006 & $-$3.042 & 0.038 &   $-$1.612 &     0.028 &   $-$3.741 &     0.034 &   $-$4.654 &     0.042 & $-$2.506 &  0.041 &    $-$1.890 &      0.054 &    $-$3.644 &      0.030 &    $-$4.396 &      0.024 \\
              & 0.002 & $-$3.161 & 0.032 &   $-$1.582 &     0.019 &   $-$3.794 &     0.026 &   $-$4.742 &     0.034 & $-$2.747 &  0.040 &    $-$1.657 &      0.051 &    $-$3.580 &      0.029 &    $-$4.404 &      0.024 \\
 \hline
              & 0.014 & $-$2.985 & 0.052 &   $-$1.785 &     0.043 &   $-$3.874 &     0.045 &   $-$4.769 &     0.053 & $-$2.409 &  0.074 &    $-$2.220 &      0.101 &    $-$3.906 &      0.056 &    $-$4.629 &      0.041 \\
        F090W & 0.006 & $-$3.113 & 0.035 &   $-$1.741 &     0.026 &   $-$3.920 &     0.031 &   $-$4.854 &     0.038 & $-$2.600 &  0.041 &    $-$1.994 &      0.054 &    $-$3.814 &      0.030 &    $-$4.594 &      0.024 \\
              & 0.002 & $-$3.219 & 0.030 &   $-$1.705 &     0.018 &   $-$3.958 &     0.025 &   $-$4.924 &     0.032 & $-$2.838 &  0.040 &    $-$1.757 &      0.050 &    $-$3.744 &      0.028 &    $-$4.595 &      0.024 \\
 \hline
              & 0.014 & $-$3.107 & 0.047 &   $-$1.997 &     0.039 &   $-$4.172 &     0.040 &   $-$5.104 &     0.048 & $-$2.606 &  0.065 &    $-$2.352 &      0.090 &    $-$4.176 &      0.049 &    $-$4.958 &      0.036 \\
        F115W & 0.006 & $-$3.215 & 0.032 &   $-$1.951 &     0.023 &   $-$4.201 &     0.028 &   $-$5.165 &     0.035 & $-$2.758 &  0.037 &    $-$2.147 &      0.048 &    $-$4.078 &      0.027 &    $-$4.905 &      0.021 \\
              & 0.002 & $-$3.306 & 0.028 &   $-$1.909 &     0.017 &   $-$4.223 &     0.023 &   $-$5.214 &     0.030 & $-$2.961 &  0.038 &    $-$1.935 &      0.047 &    $-$4.007 &      0.027 &    $-$4.895 &      0.023 \\
 \hline 
              & 0.014 & $-$3.124 & 0.044 &   $-$2.037 &     0.037 &   $-$4.224 &     0.038 &   $-$5.161 &     0.046 & $-$2.666 &  0.062 &    $-$2.343 &      0.085 &    $-$4.209 &      0.047 &    $-$5.009 &      0.034 \\
          $J$ & 0.006 & $-$3.224 & 0.031 &   $-$1.989 &     0.023 &   $-$4.246 &     0.028 &   $-$5.214 &     0.034 & $-$2.795 &  0.035 &    $-$2.159 &      0.046 &    $-$4.116 &      0.026 &    $-$4.955 &      0.020 \\
              & 0.002 & $-$3.311 & 0.028 &   $-$1.947 &     0.017 &   $-$4.265 &     0.023 &   $-$5.258 &     0.030 & $-$2.980 &  0.037 &    $-$1.963 &      0.046 &    $-$4.049 &      0.026 &    $-$4.943 &      0.022 \\
 \hline
              & 0.014 & $-$3.249 & 0.040 &   $-$2.209 &     0.033 &   $-$4.483 &     0.034 &   $-$5.458 &     0.041 & $-$2.836 &  0.056 &    $-$2.470 &      0.077 &    $-$4.455 &      0.043 &    $-$5.306 &      0.031 \\
        F150W & 0.006 & $-$3.332 & 0.028 &   $-$2.158 &     0.020 &   $-$4.491 &     0.025 &   $-$5.491 &     0.031 & $-$2.947 &  0.033 &    $-$2.285 &      0.043 &    $-$4.347 &      0.024 &    $-$5.231 &      0.019 \\
              & 0.002 & $-$3.404 & 0.026 &   $-$2.110 &     0.016 &   $-$4.493 &     0.022 &   $-$5.514 &     0.028 & $-$3.109 &  0.036 &    $-$2.099 &      0.044 &    $-$4.276 &      0.026 &    $-$5.208 &      0.021 \\
 \hline
              & 0.014 & $-$3.316 & 0.037 &   $-$2.343 &     0.031 &   $-$4.664 &     0.032 &   $-$5.658 &     0.038 & $-$2.935 &  0.053 &    $-$2.569 &      0.072 &    $-$4.624 &      0.040 &    $-$5.504 &      0.029 \\
          $H$ & 0.006 & $-$3.389 & 0.026 &   $-$2.290 &     0.019 &   $-$4.662 &     0.023 &   $-$5.679 &     0.029 & $-$3.031 &  0.032 &    $-$2.388 &      0.042 &    $-$4.509 &      0.023 &    $-$5.419 &      0.018 \\
              & 0.002 & $-$3.452 & 0.025 &   $-$2.238 &     0.015 &   $-$4.654 &     0.021 &   $-$5.689 &     0.027 & $-$3.178 &  0.035 &    $-$2.212 &      0.044 &    $-$4.436 &      0.025 &    $-$5.390 &      0.021 \\
 \hline
              & 0.014 & $-$3.352 & 0.036 &   $-$2.398 &     0.030 &   $-$4.745 &     0.031 &   $-$5.750 &     0.037 & $-$2.992 &  0.051 &    $-$2.603 &      0.070 &    $-$4.697 &      0.038 &    $-$5.595 &      0.028 \\
        $K_s$ & 0.006 & $-$3.419 & 0.026 &   $-$2.345 &     0.019 &   $-$4.738 &     0.023 &   $-$5.764 &     0.028 & $-$3.076 &  0.031 &    $-$2.428 &      0.041 &    $-$4.581 &      0.023 &    $-$5.504 &      0.018 \\
              & 0.002 & $-$3.478 & 0.025 &   $-$2.291 &     0.015 &   $-$4.726 &     0.020 &   $-$5.769 &     0.026 & $-$3.213 &  0.034 &    $-$2.259 &      0.043 &    $-$4.508 &      0.025 &    $-$5.472 &      0.021 \\
 \hline
              & 0.014 & $-$3.342 & 0.037 &   $-$2.409 &     0.030 &   $-$4.749 &     0.032 &   $-$5.751 &     0.038 & $-$2.972 &  0.051 &    $-$2.625 &      0.071 &    $-$4.705 &      0.039 &    $-$5.596 &      0.029 \\
        F277W & 0.006 & $-$3.411 & 0.026 &   $-$2.356 &     0.019 &   $-$4.744 &     0.023 &   $-$5.767 &     0.029 & $-$3.060 &  0.031 &    $-$2.448 &      0.041 &    $-$4.590 &      0.023 &    $-$5.508 &      0.018 \\
              & 0.002 & $-$3.473 & 0.025 &   $-$2.303 &     0.015 &   $-$4.734 &     0.021 &   $-$5.776 &     0.027 & $-$3.200 &  0.035 &    $-$2.276 &      0.043 &    $-$4.516 &      0.025 &    $-$5.477 &      0.021 \\
 \hline
              & 0.014 & $-$3.366 & 0.034 &   $-$2.428 &     0.029 &   $-$4.784 &     0.030 &   $-$5.794 &     0.036 & $-$3.029 &  0.049 &    $-$2.607 &      0.067 &    $-$4.727 &      0.037 &    $-$5.636 &      0.027 \\
        F356W & 0.006 & $-$3.429 & 0.025 &   $-$2.374 &     0.018 &   $-$4.775 &     0.022 &   $-$5.803 &     0.027 & $-$3.103 &  0.031 &    $-$2.441 &      0.040 &    $-$4.613 &      0.023 &    $-$5.544 &      0.018 \\
              & 0.002 & $-$3.486 & 0.025 &   $-$2.320 &     0.015 &   $-$4.760 &     0.020 &   $-$5.806 &     0.026 & $-$3.231 &  0.034 &    $-$2.280 &      0.043 &    $-$4.542 &      0.024 &    $-$5.511 &      0.021 \\
 \hline
              & 0.014 & $-$3.336 & 0.036 &   $-$2.410 &     0.030 &   $-$4.746 &     0.031 &   $-$5.746 &     0.037 & $-$2.980 &  0.051 &    $-$2.609 &      0.070 &    $-$4.695 &      0.038 &    $-$5.589 &      0.028 \\
        F444W & 0.006 & $-$3.404 & 0.026 &   $-$2.358 &     0.019 &   $-$4.740 &     0.023 &   $-$5.762 &     0.028 & $-$3.062 &  0.031 &    $-$2.440 &      0.041 &    $-$4.584 &      0.023 &    $-$5.502 &      0.018 \\
              & 0.002 & $-$3.465 & 0.025 &   $-$2.305 &     0.015 &   $-$4.730 &     0.021 &   $-$5.770 &     0.027 & $-$3.197 &  0.034 &    $-$2.274 &      0.043 &    $-$4.513 &      0.025 &    $-$5.472 &      0.021 \\
 \hline \hline
              & 0.014 & $-$3.333 & 0.038 &   $-$2.495 &     0.031 &   $-$4.828 &     0.032 &   $-$5.828 &     0.039 & $-$3.021 &  0.053 &    $-$2.641 &      0.073 &    $-$4.756 &      0.040 &    $-$5.662 &      0.030 \\
  $W_{\rm G}$ & 0.006 & $-$3.399 & 0.030 &   $-$2.446 &     0.022 &   $-$4.825 &     0.027 &   $-$5.845 &     0.033 & $-$3.054 &  0.033 &    $-$2.535 &      0.043 &    $-$4.673 &      0.024 &    $-$5.589 &      0.019 \\
              & 0.002 & $-$3.471 & 0.028 &   $-$2.399 &     0.017 &   $-$4.828 &     0.023 &   $-$5.869 &     0.030 & $-$3.162 &  0.034 &    $-$2.398 &      0.043 &    $-$4.612 &      0.024 &    $-$5.560 &      0.021 \\
 \hline
              & 0.014 & $-$3.321 & 0.038 &   $-$2.520 &     0.032 &   $-$4.845 &     0.033 &   $-$5.841 &     0.039 & $-$3.000 &  0.054 &    $-$2.676 &      0.074 &    $-$4.776 &      0.040 &    $-$5.676 &      0.030 \\
 $W_{\rm VI}$ & 0.006 & $-$3.390 & 0.031 &   $-$2.470 &     0.022 &   $-$4.844 &     0.027 &   $-$5.861 &     0.033 & $-$3.039 &  0.033 &    $-$2.566 &      0.043 &    $-$4.693 &      0.024 &    $-$5.605 &      0.019 \\
              & 0.002 & $-$3.463 & 0.028 &   $-$2.424 &     0.017 &   $-$4.848 &     0.023 &   $-$5.887 &     0.030 & $-$3.151 &  0.034 &    $-$2.425 &      0.043 &    $-$4.631 &      0.025 &    $-$5.576 &      0.021 \\
 \hline
              & 0.014 & $-$3.418 & 0.033 &   $-$2.589 &     0.027 &   $-$4.982 &     0.028 &   $-$6.007 &     0.034 & $-$3.098 &  0.047 &    $-$2.751 &      0.064 &    $-$4.919 &      0.035 &    $-$5.849 &      0.026 \\
  $W_{\rm H}$ & 0.006 & $-$3.474 & 0.024 &   $-$2.534 &     0.017 &   $-$4.966 &     0.021 &   $-$6.008 &     0.026 & $-$3.162 &  0.030 &    $-$2.586 &      0.040 &    $-$4.799 &      0.022 &    $-$5.748 &      0.017 \\
              & 0.002 & $-$3.524 & 0.024 &   $-$2.477 &     0.014 &   $-$4.944 &     0.020 &   $-$6.001 &     0.025 & $-$3.281 &  0.034 &    $-$2.429 &      0.042 &    $-$4.725 &      0.024 &    $-$5.709 &      0.020 \\
 \hline
              & 0.014 & $-$3.447 & 0.032 &   $-$2.572 &     0.027 &   $-$4.985 &     0.028 &   $-$6.019 &     0.033 & $-$3.139 &  0.046 &    $-$2.721 &      0.063 &    $-$4.919 &      0.035 &    $-$5.860 &      0.025 \\
 $W_{\rm VK}$ & 0.006 & $-$3.499 & 0.024 &   $-$2.516 &     0.017 &   $-$4.966 &     0.021 &   $-$6.015 &     0.026 & $-$3.197 &  0.030 &    $-$2.558 &      0.039 &    $-$4.796 &      0.022 &    $-$5.756 &      0.017 \\
              & 0.002 & $-$3.546 & 0.024 &   $-$2.458 &     0.014 &   $-$4.941 &     0.020 &   $-$6.005 &     0.025 & $-$3.309 &  0.033 &    $-$2.406 &      0.042 &    $-$4.722 &      0.024 &    $-$5.715 &      0.020 \\
 \hline
              & 0.014 & $-$3.519 & 0.030 &   $-$2.664 &     0.025 &   $-$5.127 &     0.026 &   $-$6.183 &     0.031 & $-$3.231 &  0.043 &    $-$2.794 &      0.060 &    $-$5.056 &      0.033 &    $-$6.025 &      0.024 \\
 $W_{\rm JK}$ & 0.006 & $-$3.562 & 0.022 &   $-$2.606 &     0.016 &   $-$5.099 &     0.019 &   $-$6.168 &     0.024 & $-$3.282 &  0.029 &    $-$2.625 &      0.039 &    $-$4.923 &      0.022 &    $-$5.908 &      0.017 \\
              & 0.002 & $-$3.601 & 0.023 &   $-$2.544 &     0.014 &   $-$5.065 &     0.019 &   $-$6.145 &     0.024 & $-$3.385 &  0.033 &    $-$2.477 &      0.041 &    $-$4.846 &      0.024 &    $-$5.861 &      0.020 \\
 \bottomrule
    \end{tabular}
\end{sidewaystable*}

\begin{table*}[h!]
    \caption{LL intercepts for the midline between the blue and red edges of the IS, for different pivot periods: $\log P_{\rm 0} = 0.0$, 0.7, and 1.0.}
    \label{table:PLR_ISedges_mid}
    \centering
    \small
    \begin{tabular}{c|c|cccccc}
\toprule
Band &  $Z$ &  $\alpha_{\rm blue+red}$ &  $\sigma_{\alpha_{\rm blue+red}}$ &  $\alpha_{\rm blue+red}$ &  $\sigma_{\alpha_{\rm blue+red}}$ &  $\alpha_{\rm blue+red}$ &  $\sigma_{\alpha_{\rm blue+red}}$ \\
              &       &  [0.0]   &  [0.0]    &  [0.7]   &  [0.7]    & [1.0]    & [1.0]     \\
\midrule
              & 0.014 &   $-$1.196 &     0.040 &   $-$3.063 &     0.033 &   $-$3.799 &     0.030 \\
 $G_{\rm BP}$ & 0.006 &   $-$1.242 &     0.036 &   $-$3.138 &     0.027 &   $-$3.863 &     0.020 \\
              & 0.002 &   $-$1.114 &     0.021 &   $-$3.056 &     0.020 &   $-$3.861 &     0.017 \\
 \hline
              & 0.014 &   $-$1.339 &     0.039 &   $-$3.238 &     0.032 &   $-$3.987 &     0.029 \\
          $V$ & 0.006 &   $-$1.383 &     0.035 &   $-$3.307 &     0.027 &   $-$4.044 &     0.020 \\
              & 0.002 &   $-$1.253 &     0.021 &   $-$3.222 &     0.020 &   $-$4.039 &     0.016 \\
 \hline
              & 0.014 &   $-$1.439 &     0.036 &   $-$3.398 &     0.029 &   $-$4.168 &     0.026 \\
          $G$ & 0.006 &   $-$1.480 &     0.034 &   $-$3.442 &     0.024 &   $-$4.213 &     0.018 \\
              & 0.002 &   $-$1.348 &     0.021 &   $-$3.360 &     0.019 &   $-$4.201 &     0.016 \\
 \hline
              & 0.014 &   $-$1.819 &     0.034 &   $-$3.878 &     0.027 &   $-$4.689 &     0.024 \\
 $G_{\rm RP}$ & 0.006 &   $-$1.853 &     0.033 &   $-$3.898 &     0.023 &   $-$4.713 &     0.017 \\
              & 0.002 &   $-$1.714 &     0.021 &   $-$3.809 &     0.018 &   $-$4.690 &     0.015 \\
 \hline
              & 0.014 &   $-$1.889 &     0.033 &   $-$3.962 &     0.027 &   $-$4.779 &     0.023 \\
          $I$ & 0.006 &   $-$1.922 &     0.033 &   $-$3.978 &     0.023 &   $-$4.800 &     0.017 \\
              & 0.002 &   $-$1.782 &     0.021 &   $-$3.888 &     0.018 &   $-$4.775 &     0.015 \\
 \hline
              & 0.014 &   $-$1.993 &     0.032 &   $-$4.114 &     0.025 &   $-$4.952 &     0.023 \\
        F090W & 0.006 &   $-$2.023 &     0.032 &   $-$4.116 &     0.022 &   $-$4.959 &     0.017 \\
              & 0.002 &   $-$1.885 &     0.021 &   $-$4.024 &     0.018 &   $-$4.929 &     0.015 \\
 \hline
             & 0.014 &   $-$2.172 &     0.029 &   $-$4.364 &     0.023 &   $-$5.230 &     0.020 \\
       F115W & 0.006 &   $-$2.199 &     0.032 &   $-$4.344 &     0.021 &   $-$5.221 &     0.016 \\
             & 0.002 &   $-$2.058 &     0.020 &   $-$4.250 &     0.017 &   $-$5.183 &     0.014 \\
 \hline
              & 0.014 &   $-$2.210 &     0.029 &   $-$4.401 &     0.026 &   $-$5.274 &     0.019 \\
          $J$ & 0.006 &   $-$2.236 &     0.031 &   $-$4.381 &     0.021 &   $-$5.264 &     0.016 \\
              & 0.002 &   $-$2.094 &     0.020 &   $-$4.288 &     0.017 &   $-$5.224 &     0.014 \\
 \hline
              & 0.014 &   $-$2.345 &     0.027 &   $-$4.604 &     0.024 &   $-$5.517 &     0.018 \\
        F150W & 0.006 &   $-$2.367 &     0.031 &   $-$4.571 &     0.020 &   $-$5.489 &     0.015 \\
              & 0.002 &   $-$2.219 &     0.029 &   $-$4.475 &     0.017 &   $-$5.442 &     0.014 \\
 \hline
              & 0.014 &   $-$2.460 &     0.026 &   $-$4.752 &     0.023 &   $-$5.686 &     0.017 \\
          $H$ & 0.006 &   $-$2.479 &     0.031 &   $-$4.713 &     0.020 &   $-$5.649 &     0.015 \\
              & 0.002 &   $-$2.323 &     0.029 &   $-$4.615 &     0.016 &   $-$5.598 &     0.014 \\
 \hline
              & 0.014 &   $-$2.506 &     0.025 &   $-$4.816 &     0.022 &   $-$5.761 &     0.016 \\
        $K_s$ & 0.006 &   $-$2.524 &     0.031 &   $-$4.774 &     0.020 &   $-$5.720 &     0.015 \\
              & 0.002 &   $-$2.364 &     0.029 &   $-$4.676 &     0.016 &   $-$5.667 &     0.014 \\
 \hline
              & 0.014 &   $-$2.520 &     0.025 &   $-$4.826 &     0.022 &   $-$5.768 &     0.016 \\
        F277W & 0.006 &   $-$2.538 &     0.031 &   $-$4.785 &     0.020 &   $-$5.728 &     0.015 \\
              & 0.002 &   $-$2.380 &     0.029 &   $-$4.687 &     0.016 &   $-$5.675 &     0.014 \\
 \hline
              & 0.014 &   $-$2.533 &     0.025 &   $-$4.845 &     0.021 &   $-$5.797 &     0.016 \\
        F356W & 0.006 &   $-$2.550 &     0.031 &   $-$4.803 &     0.020 &   $-$5.755 &     0.015 \\
              & 0.002 &   $-$2.386 &     0.029 &   $-$4.706 &     0.016 &   $-$5.701 &     0.014 \\
 \hline
              & 0.014 &   $-$2.524 &     0.025 &   $-$4.822 &     0.022 &   $-$5.764 &     0.016 \\
        F444W & 0.006 &   $-$2.543 &     0.031 &   $-$4.783 &     0.020 &   $-$5.726 &     0.015 \\
              & 0.002 &   $-$2.383 &     0.029 &   $-$4.686 &     0.016 &   $-$5.673 &     0.014 \\
 \hline \hline
              & 0.014 &   $-$2.623 &     0.026 &   $-$4.910 &     0.023 &   $-$5.859 &     0.017 \\
  $W_{\rm G}$ & 0.006 &   $-$2.641 &     0.031 &   $-$4.886 &     0.020 &   $-$5.829 &     0.015 \\
              & 0.002 &   $-$2.488 &     0.029 &   $-$4.789 &     0.016 &   $-$5.776 &     0.014 \\
 \hline
              & 0.014 &   $-$2.651 &     0.027 &   $-$4.932 &     0.023 &   $-$5.877 &     0.017 \\
 $W_{\rm VI}$ & 0.006 &   $-$2.669 &     0.031 &   $-$4.909 &     0.020 &   $-$5.849 &     0.015 \\
              & 0.002 &   $-$2.517 &     0.021 &   $-$4.812 &     0.016 &   $-$5.796 &     0.014 \\
 \hline
              & 0.014 &   $-$2.679 &     0.024 &   $-$5.019 &     0.021 &   $-$5.987 &     0.015 \\
  $W_{\rm H}$ & 0.006 &   $-$2.695 &     0.031 &   $-$4.972 &     0.020 &   $-$5.937 &     0.015 \\
              & 0.002 &   $-$2.525 &     0.028 &   $-$4.873 &     0.016 &   $-$5.879 &     0.014 \\
 \hline
              & 0.014 &   $-$2.654 &     0.024 &   $-$5.009 &     0.020 &   $-$5.987 &     0.015 \\
 $W_{\rm VK}$ & 0.006 &   $-$2.669 &     0.031 &   $-$4.960 &     0.020 &   $-$5.933 &     0.015 \\
              & 0.002 &   $-$2.496 &     0.028 &   $-$4.860 &     0.016 &   $-$5.874 &     0.014 \\
 \hline
              & 0.014 &   $-$2.724 &     0.023 &   $-$5.120 &     0.019 &   $-$6.120 &     0.014 \\
 $W_{\rm JK}$ & 0.006 &   $-$2.736 &     0.031 &   $-$5.062 &     0.020 &   $-$6.055 &     0.015 \\
              & 0.002 &   $-$2.556 &     0.028 &   $-$4.961 &     0.016 &   $-$5.992 &     0.014 \\
 \bottomrule
    \end{tabular}
    \tablefoot{The blue and red intercepts are averaged such that either the first crossing or the second+third crossings red edge is considered, depending on where the transition between the two occurs and whether the pivot period is located before or after that transition.}
\end{table*}

\clearpage

\section{Intercept-metallicity dependence for IS edges}

\begin{table}[h!]
    \caption{Metallicity parameter $\alpha_{\mathrm{M}}$ predicted at the midpoint between the IS edges, for different pivot periods: $\log P_{\rm 0} = 0.0$, 0.7, and 1.0.}
    \label{table:gamma_IS}
    \centering
    \small
    \begin{tabular}{c|cccccc}
\toprule
Band &  $\alpha_{\mathrm{M}}$ &  $\sigma_{\rm \alpha_{\mathrm{M}}}$ &  $\alpha_{\mathrm{M}}$ &   $\sigma_{\rm \alpha_{\mathrm{M}}}$ &  $\alpha_{\mathrm{M}}$ &   $\sigma_{\rm \alpha_{\mathrm{M}}}$ \\
     &  [0.0]  &  [0.0]   &  [0.7]  &  [0.7]   & [1.0]   &    [1.0] \\
\midrule
 $G_{\rm BP}$ &  $-$0.132 &    0.095 &  $-$0.043 &    0.097 &   0.051 &    0.050 \\
   $V$ &  $-$0.138 &    0.095 &  $-$0.052 &    0.096 &   0.040 &    0.047 \\
   $G$ &  $-$0.138 &    0.091 &  $-$0.071 &    0.078 &   0.023 &    0.042 \\
 $G_{\rm RP}$ &  $-$0.151 &    0.091 &  $-$0.102 &    0.064 &  $-$0.011 &    0.033 \\
   $I$ &  $-$0.152 &    0.090 &  $-$0.106 &    0.063 &  $-$0.016 &    0.032 \\
F090W &  $-$0.151 &    0.088 &  $-$0.122 &    0.054 &  $-$0.036 &    0.023 \\
F115W &  $-$0.156 &    0.085 &  $-$0.144 &    0.038 &  $-$0.060 &    0.016 \\
   $J$ &  $-$0.156 &    0.084 &  $-$0.147 &    0.038 &  $-$0.063 &    0.016 \\
F150W &  $-$0.152 &    0.098 &  $-$0.161 &    0.030 &  $-$0.089 &    0.007 \\
   $H$ &  $-$0.163 &    0.099 &  $-$0.168 &    0.026 &  $-$0.105 &    0.002 \\
  $K_s$ &  $-$0.168 &    0.099 &  $-$0.171 &    0.024 &  $-$0.112 &    0.001 \\
F277W &  $-$0.166 &    0.100 &  $-$0.171 &    0.025 &  $-$0.110 &    0.001 \\
F356W &  $-$0.173 &    0.101 &  $-$0.170 &    0.024 &  $-$0.114 &    0.001 \\
F444W &  $-$0.167 &    0.101 &  $-$0.167 &    0.026 &  $-$0.108 &    0.002 \\
  \hline
  $W_{\rm G}$ &  $-$0.161 &    0.095 &  $-$0.152 &    0.036 &  $-$0.100 &    0.009 \\
 $W_{\rm VI}$ &  $-$0.172 &    0.083 &  $-$0.152 &    0.037 &  $-$0.098 &    0.010 \\
  $W_{\rm H}$ &  $-$0.180 &    0.102 &  $-$0.177 &    0.021 &  $-$0.127 &    0.004 \\
 $W_{\rm VK}$ &  $-$0.185 &    0.102 &  $-$0.180 &    0.020 &  $-$0.133 &    0.006 \\
 $W_{\rm JK}$ &  $-$0.195 &    0.104 &  $-$0.191 &    0.014 &  $-$0.150 &    0.011 \\
 \bottomrule
\end{tabular}
\end{table}

\end{appendix}

\end{document}